\documentclass[12pt, draftclsnofoot, onecolumn]{IEEEtran}

\usepackage{cite}
\usepackage{amsmath,amssymb,amsfonts}
\usepackage{algorithm}  
\usepackage{algpseudocode}  
\usepackage{graphicx}
\usepackage{epstopdf}
\usepackage{textcomp}
\usepackage{xcolor}
\usepackage{bm}
\usepackage{enumerate}
\usepackage{subfig}
\usepackage{epsfig}
\usepackage{caption}
\ifCLASSINFOpdf
\else
\fi
\begin{document}
	
%
\title{Intelligent Omni Surfaces assisted Integrated Multi-Target Sensing and Multi-User MIMO Communications}
%
%
%
\author{Ziheng~Zhang,~Wen~Chen,~\IEEEmembership{Senior~Member,~IEEE},\\~Qingqing~Wu,~\IEEEmembership{Senior~Member,~IEEE},~Zhendong~Li,\\~Xusheng~Zhu,~and~Jinhong~Yuan,~\IEEEmembership{Fellow,~IEEE\vspace{-2em}}
\thanks{Z. Zhang, W. Chen, Q. Wu, Z. Li and X. Zhu are with the Department of Electronic Engineering, Shanghai Jiao Tong University, Shanghai 200240, China (e-mail: zhangziheng@sjtu.edu.cn; wenchen@sjtu.edu.cn; lizhendong@sjtu.edu.cn;
xushengzhu@sjtu.edu.cn;
qingqingwu@sjtu.edu.cn).}
\thanks{Jinhong Yuan is with the School of Electrical Engineering and Telecommunications, University of New South Wales, Sydney, NSW 2025, Australia
(e-mail: j.yuan@unsw.edu.au).}


\thanks{(\emph{Corresponding author: Wen Chen.})}}
\maketitle
\begin{abstract}
Drawing inspiration from the advantages of intelligent reflecting surfaces (IRS) in wireless networks, this paper presents a novel design for intelligent omni surface (IOS) enabled integrated sensing and communications (ISAC). By harnessing the power of multi-antennas and a multitude of elements, the dual-function base station (BS) and IOS collaborate to realize joint active and passive beamforming, enabling seamless 360-degree ISAC coverage. The objective is to maximize the minimum signal-to-interference-plus-noise ratio (SINR) of multi-target sensing, while ensuring the multi-user multi-stream communications. To achieve this, a comprehensive optimization approach is employed, encompassing the design of radar receive vector, transmit beamforming matrix, and IOS transmissive and reflective coefficients. Due to the non-convex nature of the formulated problem, an auxiliary variable is introduced to transform it into a more tractable form. Consequently, the problem is decomposed into three sub-problems based on the block coordinate descent algorithm. Semidefinite relaxation and successive convex approximation methods are leveraged to convert the sub-problem into a convex problem, while the iterative rank minimization algorithm and penalty function method ensure the equivalence. Furthermore, the scenario is extended to mode switching and time switching protocols. Simulation results validate the convergence and superior performance of the proposed algorithm compared to other benchmark algorithms.

\end{abstract}

\begin{IEEEkeywords}
Intelligent omni
surface, 
integrated sensing and communication, 
multi-stream communication, 
multi-target sensing.

\end{IEEEkeywords}

%
\IEEEpeerreviewmaketitle

\section{Introduction}
The next-generation wireless network is poised to enable intelligent and diverse scenarios, necessitating the development of mobile network systems with multidimensional capabilities. In addition to communications, sensing and computation play crucial roles in future networks, with communication and sensing systems sharing many commonalities in signal processing and radio frequency (RF) units\cite{10077112}. Specifically, higher frequency signals with wide bandwidth enhance transmission rates and sensing resolution, while larger-scale antenna arrays improve beam accuracy. Furthermore, wireless communication and sensing systems exhibit similarities in waveform design, signal processing and resource allocation \cite{9737357}. Therefore, it is highly meaningful to design a software and hardware resource-sharing system that enables integrated sensing and communication (ISAC). This approach promises to enhance spectral efficiency and reduce hardware costs in next-generation wireless networks.

The essence of ISAC lies in striking a balance between communication and sensing systems, and it is important to identify appropriate performance metrics that reveal this fundamental trade-off. 
Communication rate and Fisher information are widely recognized as classical performance metrics for communication and radar systems, respectively \cite{7279172}. To characterize this trade-off, the notion of the Cramér-Rao bound-rate region has been introduced. The points within this region portray the performance of the ISAC system, thereby serving as a guiding framework for the design of the actual system \cite{10001144}. However, designing a suitable unified waveform to balance the trade-off in deterministic-random is challenging, as randomness can convey information but may reduce sensing performance. A natural type of approach is  embedding information onto radar signals.
Chirp signals, extensively employed in radar systems, possess modulable parameters such as frequency, phase, and amplitude, making them suitable for shift keying \cite{4268440}. Another method is transmitting data during the waiting time of pulsed radars, thereby achieving information embedding in the time domain. This method reduces mutual interference between the transmitted signal and the echo signal, enabling full duplex operation 
\cite{9724187}. However, the performance of this method can be affected by the distance and number of clutters. To mitigate environmental interference, an alternative method of conveying information involves embedding data into the spatial domain. In this approach, the main beam of the radar system is employed for sensing tasks, while the sidelobe level is harnessed to transmit the encoded bitstream \cite{7347464}. Additionally, improving the sensing ability of communication signals is another type of design approach. 
Orthogonal frequency division multiplexing-multiple input multiple out
(OFDM-MIMO) signals are designed to achieve multi-beam to sensing multi-target. Intercarrier interference can be exploited to extract the target's speed information \cite{9529026}. Recently, an orthogonal delay-Doppler division multiplexing (ODDM) or generally delay-Doppler domain multicarrier modulation (DDMC) has been proposed as a potential new waveform for future ISAC applications. Its design and performance in practical ISAC systems are not yet reported \cite{9829188,9838684}. Although a fully unified waveform improves resource utilization, its implementation requires more complex signal processing and hardware support. 

Non-overlapped waveform design is an additional method employed to achieve ISAC, effectively mitigating interference between radar and communication systems. Analogous to multiple access methods in communication, resource allocation for ISAC can be realized in domains such as time, frequency, spatial, and code. In \cite{9728752}, a time-division ISAC protocol was proposed for the Internet of vehicles. The protocol adopts a subframe structure comprising both communication and sensing symbols, enabling the efficient and timely sharing of sensing data. Moreover, resource allocation can also be extended to the frequency domain, as discussed in \cite{8094973}. To further enhance the utilization of wireless resources, Shi \emph{et al.} proposed a successive interference cancellation signal processing scheme, which enables code division in an OFDM system. Notably, this scheme aims to improve the performance of the system by mitigating interference. Additionally, the authors established a unified channel model that comprehensively incorporates both communication and sensing aspects. The model integrates essential parameters such as Doppler shift, radar echo model, fading coefficients, average power, and delay to effectively characterize the ISAC channel properties \cite{9359665}. With the advancements in large-scale antenna arrays, MIMO technology is increasingly employed in communication and sensing systems. Consequently, spatial division to maximize multiplexing gain has gained popularity in ISAC. Liu \emph{et al.} minimized the difference between the optimal sensing beam and the ISAC beam that satisfies the SINR requirement of multi-user communication using manifold algorithms \cite{8288677}. To reduce the complexity of spatial division, designing the covariance matrix of the transmitted signal can simplify the precoding problem, which transforms the requirements for transmission power, sensing and communication into convex constraints. To further reduce interference between multi-user, this work in \cite{9724205} introduced dirty paper coding. Simulation results show that ISAC can achieve multi-user communication with only 1dB radar sensing beam loss compared with traditional radar design. The ISAC research directions mentioned above have mainly focused on scenarios with multiple-input single-output (MISO) or single-target sensing. However, these approaches may not be well-suited for current communication or sensing systems as they lack the necessary adaptability. In addition, ISAC systems often exhibit a heavy reliance on line-of-sight signals, which results in limited coverage of the surrounding building structures, particularly in urban environments. Therefore, further research is necessary to enhance the system's performance and improve model universality.

Intelligent reflecting surface (IRS) is a potential technology to break the bottleneck
in the development of ISAC. As a passive communication unit, IRS enables channel reconstruction and beamforming to enhance signals and overcome line-of-sight obstructions, and it has found extensive applications in wireless communication \cite{9086766}. Based on these characteristics, IRS is compatible with radar systems. In recent studies \cite{9361184,9456027}, passive IRS has been investigated as an assistant to mono-static radar systems where the receiving and transmitting antennas are colocated. In \cite{9361184}, IRS is employed to assist distributed radar, improving the detection probability of targets compared to traditional systems. Zhang \emph{et al.} enhanced signal strength in multi-user localization with an IRS-assisted system by optimizing the decision function and IRS coefficients, establishing the relationship between system performance and the number of elements \cite{9456027}. However, the utilization of IRS in mono-static radar systems may lead to an increase in path loss compared to conventional systems. In order to tackle this challenge, a novel IRS-assisted sensing architecture is proposed, capitalizing on the active IRS's ability to transmit and receive signals \cite{9724202}. Similarly, in \cite{9938373}, the authors presented a method that deploys a limited number of sensors within the IRS to reduce the adverse impact caused by excessive reflections. It is important to emphasize that the aforementioned architecture requires the IRS to possess signal processing capabilities, leading to an inevitable increase in the deployment cost of the IRS. To control the deployment cost of IRS and reduce path loss in the sensing system, bi-static radar is a feasible design. Deploying multiple antennas in a distributed manner can mitigate the effects of path loss caused by reflection from the IRS, while also avoiding self interferences of radar signals at BS \cite{9705498}.   

IRS has gained significant attention in both communication and sensing domains, leading to ongoing research on IRS-assisted ISAC \cite{9416177,9729741,10131933,10143420}. In \cite{9416177}, an approach utilizing IRS is proposed to mitigate interference among multi-user while maintaining similarity with sensing-only beams. The authors further investigate the performance improvements and trade-offs in ISAC compared to systems without IRS. Due to the ability of multiple IRS to achieve more precise beamforming, He \emph{et al.} proposed a double-IRS-assisted ISAC system model where the communication and sensing systems operate with separate transmission antennas \cite{9729741}. Two IRSs are strategically positioned near the transmitting and receiving antennas to eliminate interference between the two systems. However, conventional IRS is limited to signal reflection\cite{9424177} or transmission \cite{9570775} and only provides coverage in a half-space. To overcome this limitation, intelligent omni surface (IOS), also known as simultaneously transmitting and reflecting (STAR)-IRS, has emerged as a promising solution \cite{9365009,9690478}. The application of IOS in wireless communication is explored in \cite{9570143}, where Mu \emph{et al.} compared the performance of IOS in various scenarios with conventional IRSs and demonstrate that IOS exhibits superior performance in both unicast and multicast communication. Moreover, the benefits of IOS can also be extended to ISAC. Wang \emph{et al.} proposed a scheme where the space is divided into two halves: the transmitting space and the reflecting space, dedicated to communication and sensing, respectively \cite{10050406}. This approach effectively reduces inter-user interference but imposes limitations on the placement of communication users and sensing targets.

Motivated by the aforementioned backgrounds and challenges, our objective is to achieve enhanced spatial division multiplexing gain in the ISAC system, accomplish 360-degree coverage and ensure compatibility with existing communication systems. To address these goals, we propose an ISAC system empowered by IOS, wherein a dual-function base station (BS) transmits signals for multi-user communication and detects multiple targets, while the IOS reflects and simultaneously transmits incident signals. To mitigate path loss resulting from multiple signal reflections and simplify the deployment of IOS-assisted sensing systems, we introduce bistatic radar into our model. 
The main focus of this paper is to maximize the minimum signal-to-interference-plus-noise ratio (SINR) of multi-target sensing, while ensuring that the achievable rates of all users surpass a predefined threshold. Since the formulated problem is non-convex, it is imperative to devise an efficient algorithm to solve it. The main contributions of this paper can be summarized as follows:
\begin{itemize}
\item We propose an IOS-enabled ISAC system that leverages a dual-function BS and IOS with multiple antennas and numerous elements. This system implements joint active and passive beamforming to serve users and detect targets. In order to enhance practicality, we consider the impact of echo interference caused by multi-target, as well as multi-stream communication. To achieve 360-degree ISAC, we formulate a maximization problem for minimum SINR of multi-target. This problem involves joint optimization of the radar receive vectors, transmit beamforming matrices, and IOS reflective and transmissive coefficients. However, due to the high coupling of variables, both the objective function and the communication achievable rate constraint are non-convex. Consequently, solving this problem poses a significant challenge.

\item We propose a joint optimization algorithm to address the SINR maximization problem in the context of energy splitting (ES) protocol. Specifically, we decompose the non-convex problem into three sub-problems using the coordinate descent (BCD) algorithm framework.
The sub-problem is converted into a convex problem through the utilization of semidefinite relaxation and the successive convex approximation method. Additionally, the equivalence is guaranteed through the use of iterative rank minimization algorithms and the penalty function method. Furthermore, we extend the proposed algorithm to mode switching (MS) and time switching (TS) protocols, allowing for more flexibility and adaptability in different scenarios.

\item Numerical results demonstrate the convergence and effectiveness of the proposed algorithm for the ES protocol. Additionally, our results reveal that
the number of transmit/receive antennas has a more significant impact on performance compared with 
the number of IOS elements or maximum transmit power does. The relative position of each target to the BS also makes a great contribution to system performance, indicating the importance of BS location selection.
\end{itemize}

The rest of this paper is as follows. In Section II, we introduce the system model of IOS-assisted integrated multi-target sensing and multi-user MIMO communications and formulate the optimization problem. Then, in Section III, We elaborate on the joint optimization algorithm. Section IV reveals the convergence and performance superiority of the proposed algorithm compared to other benchmarks. Finally, Section V concludes this paper.

\textit{Notations:} Matrices are represented by bold uppercase letters. Vectors are denoted by bold lowercase letters. Scalars are represented by standard lowercase letters. For a complex-valued scalar $x$, $\left| {x} \right|$ denotes its absolute value. For a general matrix $\bf{A}$, $\text{rank}(\bf{A})$ and ${\bf{A}}^H$ denote its rank, conjugate transpose, respectively. For a square matrix $\bf{X}$, $\text{rank}(\bf{X})$, ${\bf{X}}^H$, and $\text{tr}(\bf{X})$ denote its rank, conjugate transpose, and trace. ${\bf{X}} \succeq 0$ denotes that $\bf{X}$ is a positive semidefinite matrix. ${\mathbb{C}^{M \times N}}$ 
represents the ${M \times N}$ dimensional complex matrix space and $j$ is the imaginary unit. 
\section{System Model And Problem Formulation}
As shown in Fig. 1, the system model of IOS-assisted integrated multi-target sensing and multi-user MIMO communications networks is first introduced. The model consists of crucial components, namely a multi-antenna base station enabling ISAC, an IOS that facilitates the transmission of ISAC signals in the reflective and transmissive areas, $K$ multi-antenna users and $O$ sensed targets. The direct links between BS and users or targets are blocked, and this is a typical application scenario of the IRS. To mitigate self-interference and minimize the attenuation effect of IOS on signals, the model of a bistatic radar is considered. Specifically, receiving antennas are implemented in both the reflective and the transmissive area to receive echo signals. The dual function BS is equipped with ${{N}_{t}}$ transmitting antennas, and the receiving arrays in two areas are quipped with ${{N}_{r}}$ receiving antennas.

\begin{figure}
\centerline{\includegraphics[width=12cm]{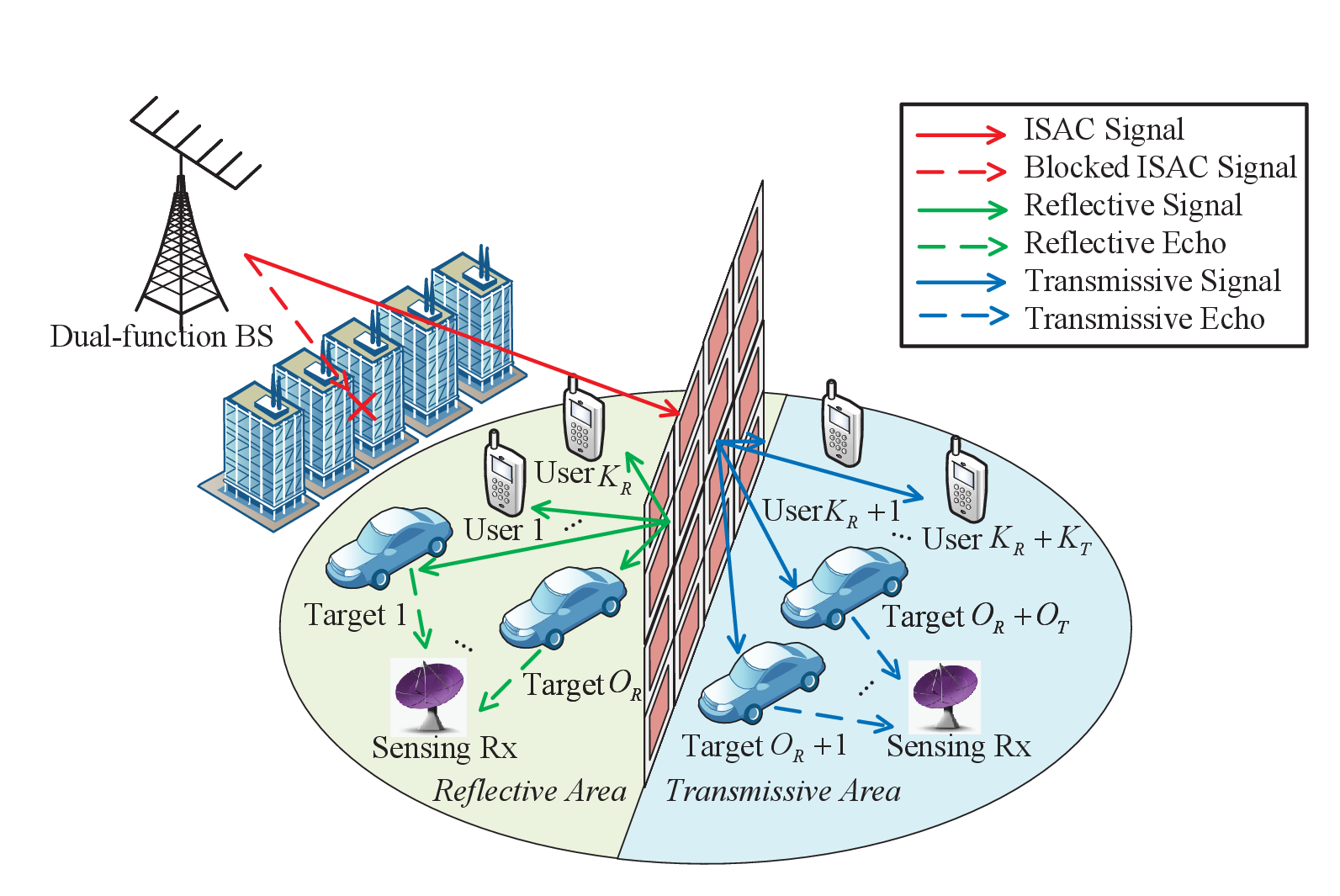}}
	\caption{IOS enables ISAC networks.}
	\label{Fig1}
\end{figure}
It is assumed that there are ${{K}_{R}}$ multi-antenna users in the coverage range of the IOS reflective signal, indexed by $k\in \left\{ 1,\cdots, {{K}_{R}} \right\}$ and ${{K}_{T}}$ multi-antenna users in the coverage range of the IOS transmissive signal, indexed by $k\in \left\{ {{K}_{r}}+1,\cdots, {{K}_{R}}+{{K}_{T}} \right\}$, where ${{K}_{T}}+{{K}_{R}}=K$. The number of receiving antennas of user $k$ is ${{M}_{k}}$. Similarly, targets that need to be sensed are also located in the reflective and transmissive areas. There are ${{O}_{R}}$ targets located in the reflection area, indexed by $i\in \left\{ 1,2,\cdots, {{O}_{R}} \right\}$ and ${{O}_{T}}$ targets located in the transmissive area, indexed by $i\in \left\{ {{O}_{R}}+1,{{O}_{R}}+2, \cdots ,{{O}_{R}}+{{O}_{T}} \right\}$, where ${{O}_{R}}+{{O}_{T}}=O$.

Considering that the communication scenario is downlink, the BS is required to transmit multi stream data with dimension ${{D}_{k}}$ to user $k$ simultaneously. In addition, in order to ensure the performance of the radar, additional signal streams need to be introduced. By introducing an additional sensing stream equal to the number of transmit antennas, the full rank of the autocorrelation matrix of the transmitted signal can be ensured \cite{9652071}. The signal transmitted by the BS can be represented as
\begin{align}
{\bf{x}} = \sum\limits_{k = 1}^K {{{\bf{W}}_k}{{\bf{d}}_k}}  + \sum\limits_{t = 1}^{{N_T}} {{{\bf{m}}_t}{s_t}},
\end{align}
where ${{\mathbf{d}}_{k}}\in {{\mathbb{C}}^{{{D}_{k}}\times 1}}$ represents the multi-stream data sent by the BS to the user $k$. In order to ensure that the data is well received, the transmitted data needs to meet ${{D}_{k}}\le \min \left\{ {{N}_{t}},{{M}_{k}} \right\}$. It is assumed that the data ${{\mathbf{d}}_{k}}$ between different users is independent and follows a Gaussian distribution ${{\mathbf{d}}_{k}}\sim \mathcal{C}\mathcal{N}\left( 0,{\mathbf{I}_{{{D}_{K}}}} \right)$. ${{\bf{W}}_k}\in {{\mathbb{C}}^{{{N}_{T}}\times {{D}_{k}}}}$ represents the communication beamforming matrix of the user $k$. 
${{s}_{t}}$ denotes the orthogonal base-band radar signal transmitted by the $t$-th antenna and ${{\bf{m}}_t}$ represents the beamforming vector of the $t$-th sensing stream. The power of the transmitted signal can then be expressed as
\begin{align}
{P_t} = E\left( {\left\| {\bf{x}} \right\|_2^2} \right) = {\rm{tr}}\left( {\sum\limits_{k = 1}^K {{{\bf{W}}_k}{\bf{W}}_k^H}  + \sum\limits_{t = 1}^{{N_T}} {{{\bf{m}}_t}{\bf{m}}_t^H} } \right).
\end{align}
\subsection{IOS Model}
The IOS system is composed of $N$ elements for ISAC, indexed by $n\in \left\{ 1,\cdots ,N \right\}$ and its reflective and transmissive coefficient matrices are respectively represented as
\begin{align}
{{\bf{\Theta }}_R} = {\text{diag}}\left( {{{\bf{\theta }}_R}} \right) = {\text{diag}}\left( {\left[ {\sqrt {{\alpha _{1,R}}} {e^{j{\beta _{1,R}}}},\sqrt {{\alpha _{2,R}}} {e^{j{\beta _{2,R}}}}, \cdots \sqrt {{\alpha _{N,R}}} {e^{j{\beta _{N,R}}}}} \right]} \right),\\
{{\bf{\Theta }}_T} = {\text{diag}}\left( {{{\bf{\theta }}_T}} \right) = {\text{diag}}\left( {\left[ {\sqrt {{\alpha _{1,T}}} {e^{j{\beta _{1,T}}}},\sqrt {{\alpha _{1,T}}} {e^{j{\beta _{2,T}}}}, \cdots \sqrt {{\alpha _{1,T}}} {e^{j{\beta _{N,T}}}}} \right]} \right),
\end{align}
where $\sqrt{{{\alpha }_{n,R}}}{{e}^{j{{\beta }_{n,R}}}}$ and $\sqrt{{{\alpha }_{n,R}}}{{e}^{j{{\beta }_{n,R}}}}$ represent the coefficients of reflective and transmissive for the $n$-th element respectively, and ${{\alpha }_{n,R}},{{\alpha }_{n,T}}\in \left[ 0,1 \right],{{e}^{j{{\beta }_{n,R}}}},{{e}^{j{{\beta }_{n,T}}}}\in \left[ 0,2\pi  \right)$. According to power constraints, the amplitude of the coefficients must satisfy the following condition
\begin{align}
{{\alpha }_{n,R}}+{{\alpha }_{n,T}}=1,\forall n.
\end{align}
There are three common protocols utilized in IOS-assisted wireless communication: energy splitting (ES), mode switching (MS), and time switching (TS). In the upcoming section, we will initially develop and address the problem based on the ES protocol. Subsequently, in sections \uppercase\expandafter{\romannumeral3}-$E$ and \uppercase\expandafter{\romannumeral3}-$F$, we will explore the modifications introduced by the MS and TS protocols, respectively, along with their respective solutions.

(1) ES: In the ES protocol, each IOS element operates in both reflection and transmission modes, dividing the incident energy into reflective and transmissive signals, with an energy ratio denoted as ${{\alpha }_{n,R}}:{{\alpha }_{n,T}}$, where ${{\alpha }_{n,R}},{{\alpha }_{n,T}}\in \left[ 0,1 \right],\forall n$. The ES protocol offers flexibility in design but introduces significant information exchange overhead.

(2) MS: In the MS protocol, each IOS element of IOS operates either in reflective mode or transmissive mode, but not both simultaneously. The $N$ IOS elements are divided into two groups: ${{N}_{R}}$ elements operate in the reflective mode and ${{N}_{T}}$ elements operate in the transmissive mode. It is important to note that ${{N}_{R}}+{{N}_{T}}=N$ and ${{\alpha }_{n,R}},{{\alpha }_{n,T}}\in \left\{ 0,1 \right\},\forall n$.
Compared with the ES protocol, the MS protocol cannot achieve the same beam gain, but it is relatively easier to implement.

(3) TS: In the TS protocol, all elements of the IOS work in the same mode periodically. The time proportion of IOS elements working in the transmission mode is denoted as $0\le {{\lambda }_{R}}\le 1$, while the time proportion in the reflection mode is denoted as $0\le {{\lambda }_{T}}\le 1$, where ${{\lambda }_{R}}+{{\lambda }_{T}}=1$. Under this protocol, the coefficient design for each IOS element is relatively simple as the transmission and reflection coefficients are not coupled. However, precise synchronization in the time domain is essential for accurate implementation of this protocol.

\subsection{Channel Model}
To represent the downlink model, we adopt the Rician channel modeling, assuming all channels as quasi-static fading. The channel from the BS to the IOS can be represented as
\begin{align}\label{hbr}
{{\bf{H}}_{{\rm{RB}}}} = \sqrt {\beta {{\left( {\frac{{{d_{{\rm{RB}}}}}}{{{d_0}}}} \right)}^{ - \alpha }}} \left( {\sqrt {\frac{\kappa }{{\kappa  + 1}}} {{\bf{H}}_{{\rm{RB}},{\rm{LoS}}}} + \sqrt {\frac{1}{{\kappa  + 1}}} {{\bf{H}}_{{\rm{RB}},{\rm{NLoS}}}}} \right),
\end{align}
where $\alpha$ represents the path loss exponent, $\beta$ represents the channel loss exponent when the reference distance
${{d}_{0}}=1$ meter (m), and ${{d}_{\text{RB}}}$ represents the distance between BS and IOS.
${{\mathbf{H}}_{\text{RB,NLoS}}}\sim \mathcal{C}\mathcal{N}\left( 0,{{\mathbf{I}}_{N\times {{N}_{t}}}} \right)$ and  ${{\bf{H}}_{{\rm{RB,LoS}}}}$ represent the LoS component and NLoS component respectively. 
Here, we assume that the IOS is a uniform planar array, i.e., $N={{N}_{x}}\times {{N}_{z}}$, ${{N}_{x}}$ and ${{N}_{z}}$ represent the number of elements in the horizontal and vertical directions of the IOS respectively. The transmitting antenna at the BS is a uniform linear array, and ${{\mathbf{H}}_{\text{RB,LoS}}}$ can be expressed as
\begin{align}
{{\mathbf{H}}_{\text{RB,LoS}}}=\mathbf{u}\left( {{\theta }_{\text{RB,h}}},{{\theta }_{\text{RB,v}}} \right)\cdot {{\mathbf{l}}^{T}}\left( {{\theta }_{\text{RB,h}}} \right)\in {{\mathbb{C}}^{N\times {{N}_{t}}}},
\end{align}
where ${\bf{l}}\left( {{\theta _{{\rm{RB,h}}}}} \right) = {\bf{a}}\left( {{\varphi _{{\rm{BS}}}},{N_t}} \right)$ denotes the transmit vector from the BS to IOS, $\mathbf{u}\left( {{\theta }_{\text{RB,h}}},{{\theta }_{\text{RB,v}}} \right)=\mathbf{a}\left( {{\varphi }_{\text{IOS,h}}},{{N}_{x}} \right)\otimes \mathbf{a}\left( {{\varphi }_{\text{IOS,v}}},{{N}_{z}} \right)$ denotes the receive vector from the BS to IOS, ${\bf{a}}\left( {\tilde \theta ,\tilde N} \right)$ represents the steering vector where $\tilde{\theta }$ represents the phase difference between two adjacent elements or antennas, and $\tilde{N}$ represents the number of antennas transmitted or received, which can be expressed as
\begin{align}
\mathbf{a}\left( \tilde{\theta },\tilde{N} \right)={{\left[ 1,{{e}^{j2\pi \tilde{\theta }}},\cdots {{e}^{j2\pi \left( \tilde{N}-1 \right)\tilde{\theta }}} \right]}^{T}}\in {{\mathbb{C}}^{\tilde{N}\times 1}}.
\end{align}
and ${{\varphi }_{\text{RB,h}}}={{{f}_{c}}{{d}_{\text{IOS,x}}}\sin \left( {{\theta }_{\text{RB,h}}}\right)\sin \left({{\theta }_{\text{RB,v}}} \right)}/{c}$, ${{\varphi }_{\text{RB,v}}}={{{f}_{c}}{{d}_{\text{IOS,z}}}\sin \left( {{\theta }_{\text{RB,h}}} \right)\cos \left( {{\theta }_{\text{RB,v}}} \right)}/{c}$ and ${{\varphi }_{\text{BS}}}={{{f}_{c}}{{d}_{\text{BS}}}\sin \left( {{\theta }_{\text{RB,h}}} \right)}/{c}$. ${{d}_\text{{IOS,x}}}$ and ${{d}_{\text{IOS,z}}}$ represent horizontal and vertical distances of IOS unit respectively. ${{\theta }_\text{{RB,h}}}$ and ${{\theta }_{\text{RB,v}}}$ represent respectively the physical azimuth and vertical angles-of-arrival (AoA) at the IOS respectively. ${{d}_{\text{BS}}}$ represents the distance between adjacent transmitting antennas, 
${{f}_{c}}$ represents carrier frequency.
Similarly, the channel from IOS to the $k$-th user can be represented as
\begin{align}\label{hru}
{{\mathbf{H}}_{k,\text{UR}}}=\sqrt{\beta {{\left( \frac{{{d}_{k\text{,UR}}}}{{{d}_{0}}} \right)}^{-\alpha }}}\left( \sqrt{\frac{\kappa }{\kappa +1}}{{\mathbf{H}}_{k\text{,UR,LoS}}}+\sqrt{\frac{1}{\kappa +1}}{{\mathbf{H}}_{k\text{,UR,NLoS}}} \right),\forall k.
\end{align}
The definition of physical quantities in Eq. (\ref{hru}) is similar to that in Eq. (\ref{hbr})

\subsection{Communication Model}
For the $k$-th user, the received signal can be represented as
\begin{equation}
\begin{split}
{{\mathbf{y}}_{k}}=\left( {{\mathbf{H}}_{k\text{,UR}}}{{\mathbf{\Theta }}_{{{S}_{K}}\left( k \right)}}{{\mathbf{H}}_{\text{RB}}} \right)\mathbf{x}+{{\mathbf{n}}_{k}},\forall k,
\end{split}
\end{equation}
where ${\mathbf{y}_{k}}\in {{\mathbb{C}}^{{{M}_{k}}\times 1}}$ represents the signal received by the $k$-th user, $\mathbf{x}\in {{\mathbb{C}}^{{{N}_{t}}\times 1}}$ represents the signal transmitted by the dual function BS,
${\mathbf{n}_{k}}$ represents additive white Gaussian noise (AWGN) introduced at the $k$-th user’s receiving antenna and
${\mathbf{n}_{k}}\sim \mathcal{C}\mathcal{N}\left( 0,\sigma _{n}^{2}{\mathbf{I}_{{{M}_{k}}}}\right)$. ${{S}_{K}}\left( k \right)$ is a character function that indicates the area where the $k$-th user is located, which can be expressed as
\begin{equation}  
{{S}_{K}}\left( k \right)=\left\{ 
\begin{aligned}
  & R,1\le k\le {{K}_{R}},\\ 
 & T,{{K}_{R}}+1\le k\le {{K}_{R}}+{{K}_{T}},\\ 
\end{aligned}\right.
\end{equation}

For MIMO communications, the achievable rate (bps/Hz) of $k$-th user can then be expressed as
\begin{align}
{{R}_{k}}=\log \det \left( {{\mathbf{I}}_{{{D}_{k}}}}+\mathbf{W}_{k}^{H}\mathbf{H}_{k}^{H}\mathbf{J}_{k}^{-1}{{\mathbf{H}}_{k}}{{\mathbf{W}}_{k}} \right),\forall k,
\end{align}
where ${{\mathbf{H}}_{k}}={{\mathbf{H}}_{k\text{,UR}}}{{\mathbf{\Theta }}_{{{S}_{K}}\left( k \right)}}{{\mathbf{H}}_{\text{RB}}}$ and ${{\mathbf{J}}_{k}}=\sigma _{k}^{2}{{\mathbf{I}}_{{{M}_{k}}}}+\sum\limits_{l=1,l\ne k}^{K}{{{\mathbf{H}}_{k}}{{\mathbf{W}}_{l}}\mathbf{W}_{l}^{H}\mathbf{H}_{k}^{H}}+\sum\limits_{t=1}^{{{N}_{T}}}{{{\mathbf{H}}_{k}}{{\mathbf{m}}_{t}}\mathbf{m}_{t}^{H}\mathbf{H}_{k}^{H}}$. The communications quality-of-service (QoS) requirements for all users must be satisfied, meaning that the achievable rate for each user ${{R}_{k}}$ should be greater than or equal to a certain threshold, ${{R}_{\text{th}}}$, i.e., ${{R}_{k}}\ge {{R}_{\text{th}}},\forall k$.

\subsection{Sensing Model}
When the receiving arrays process the echo signal of the $i$-th target, the echo signals of the other targets can be considered as interference. Thus, the received signal can be represented as
\begin{equation}
\begin{split}
  & {{\mathbf{p}}_{i}}={{\alpha }_{i}}\mathbf{A}\left( {{\theta }_{i,\text{OR,h}}},{{\theta }_{i,\text{OR,v}}},{{\theta }_{i,\text{RO,h}}} \right){{\mathbf{\Theta }}_{{{S}_{o}}\left( i \right)}}{{\mathbf{H}}_{\text{RB}}}\mathbf{x} \\ 
 & +\sum\limits_{o=1+\delta \left( {i - {O_R}} \right)\cdot{O_R},o\ne i}^{{{O}_{R}+\delta \left( {i - {O_R}} \right)\cdot{O_T}}}{{{\alpha }_{o}}\mathbf{A}\left( {{\theta }_{o,\text{OR,h}}},{{\theta }_{o,\text{OR,v}}},{{\theta }_{o,\text{RO,h}}} \right){{\mathbf{\Theta }}_{{{S}_{o}}\left( i \right)}}{{\mathbf{H}}_{\text{RB}}}\mathbf{x}}+{\mathbf{z}_{i}},\forall i,
\end{split}
\end{equation}
where $\mathbf{A}\left( {{\theta }_{i,\text{OR,h}}},{{\theta }_{i,\text{OR,v}}},{{\theta }_{i,\text{RO,h}}} \right)=\mathbf{l}\left( {{\theta }_{i,\text{RO,h}}} \right)\cdot {{\mathbf{u}}^{T}}\left( {{\theta }_{i,\text{OR,h}}},{{\theta }_{i,\text{OR,v}}} \right)\in {{\mathbb{C}}^{{{N}_{r}}\times {{N}_{t}}}}$, ${{\theta }_{i,\text{OR,h}}}$ and ${{\theta }_{i,\text{OR,v}}}$ denote the azimuth and vertical AoAs from the IOS to target,	${{\theta }_{i,\text{RO,h}}}$ denote the azimuth angles-of-departure (AoDs) from the targe to receiving antenna. ${
\mathbf{z}
_{i}}$ represents Gaussian white noise at the receiving antenna and ${{\bf{z}}_i} \sim \mathcal{C}\mathcal{N}\left( {0,\sigma _i^2{{\bf{I}}_{{N_r}}}} \right)$. $\delta \left( x \right)=1,x>0;\delta \left( x \right)=0,x\le 0$. ${{S}_{O}}\left( i \right)$ is a character function that indicates the area where the $i$-th target is located, which can be expressed as
\begin{equation}
{{S}_{O}}\left( i \right)=\left\{\begin{aligned}
 & R,1\le i\le {{O}_{R}},\\ 
 & T,{{O}_{R}}+1\le i\le {{O}_{R}}+{{O}_{T}},\\ 
\end{aligned} \right.
\end{equation}
The BS utilizes multiple antennas to achieve receiving beamforming. We use ${\bf{g}_{i}}$ to represent the filter when receiving the echo from the $i$-th target and the SINR of the echo signal of the $i$-th target can be expressed as
\begin{align}
{\text{SIN}}{{\text{R}}_i} = \frac{{\alpha _i^2{{\left| {{\bf{g}}_i^H{{\bf{S}}_i}{\bf{x}}} \right|}^2}}}{{{\bf{g}}_i^H\left( {\sum\limits_{o=1+\delta \left( {i - {O_R}} \right)\cdot{O_R},o\ne i}^{{{O}_{R}+\delta \left( {i - {O_R}} \right)\cdot{O_T}}} {\alpha _o^2{{\bf{S}}_o}{\bf{x}}{{\bf{x}}^H}{\bf{S}}_o^H}  + \sigma _i^2{\bf{I}}_{{{N}_{r}}}} \right){{\bf{g}}_i}}},\forall i.
\end{align}
where ${{\mathbf{S}}_{i}}=\mathbf{A}\left( {{\theta }_{i,\text{OR,h}}},{{\theta }_{i,\text{OR,v}}},{{\theta }_{i,\text{RO,h}}} \right){{\mathbf{\Theta }}_{{{S}_{o}}\left( i \right)}}{{\mathbf{H}}_{\text{RB}}}$ and it represents the channel from the BS to the radar receive array for the echo signal from the $i$-th target.

\subsection{Problem Formulation}
In multi-target detection, it is necessary to ensure the echo signal of all targets can be effectively detected. Because the detection performance of the radar is correlated with the SINR of the perceived signal, the goal of the system design is to maximize the SINR of the multi-target. Therefore, the design problem of the IOS-assisted ISAC system can be expressed as
\begin{subequations}\label{p0}
\begin{align}
 \text{(P0)}:\quad &\underset{\begin{smallmatrix}
 {{\mathbf{W}}_{k}},{{\mathbf{m}}_{t}},{{\mathbf{g}}_{i}} \\ 
{{\theta }_{R}},{{\theta }_{T}} 
\end{smallmatrix}}{\mathop{\max}}\,\underset{i}{\mathop{\min}}\quad\text{SIN}{{\text{R}}_{i}}, \\ 
\rm{s.t.}\quad &\text{tr}\left( \sum\limits_{k=1}^{K}{{{\mathbf{W}}_{k}}\mathbf{W}_{k}^{H}}+\sum\limits_{t=1}^{{{N}_{T}}}{{{\mathbf{m}}_{t}}\mathbf{m}_{t}^{H}} \right)\le {{P}_{\max }}, \\ 
&{{R}_{k}}\ge {{R}_{\text{th}}},\forall k, \\ 
&{{\alpha }_{n,R}},{{\alpha }_{n,T}}\in \left[ 0,1 \right],{{\alpha }_{n,R}}+{{\alpha }_{n,T}}=1,\forall n, \\ 
&{{\beta }_{n,R}},{{\beta }_{n,T}}\in \left[ 0,2\pi\right),\forall n.
\end{align}
\end{subequations}
where $P_{\max }$ indicates the maximum transmission power of the BS. (\ref{p0}b) denotes the BS transmit power limitation.
(\ref{p0}c) guarantees the performance of multi -user multi-stream communications.
(\ref{p0}d) and (\ref{p0}e) represent the constraints on the design of IOS amplitude and phase coefficients respectively.
Because the objective function and constraints in problem (P0) are non-convex and the variables therein are highly coupled, it is difficult to directly solve the primal problem and we propose a joint active and passive beamforming design algorithm in the following.  
\section{Joint Active And Passive Beamforming Design For IOS-Assisted ISAC Network}
We divide the problem (P0) into three sub-problems to solve based on the BCD algorithm. Specifically, the sub-problem 1 is to optimize radar receive beamforming vectors ${\mathbf{g}_{i}}$ for fixed IOS reflective and transmissive coefficient ${{\mathbf{\theta }}_{R}},{{\mathbf{\theta }}_{T}}$ and beamforming matrices ${\mathbf{W}_{k}},{\mathbf{m}_{t}}$ . In the sub-problem 2, the radar receive beamforming vectors ${\mathbf{g}_{i}}$ and IOS reflective and transmissive coefficient ${{\mathbf{\theta }}_{R}},{{\mathbf{\theta }}_{T}}$ are fixed, beamforming matrices ${\mathbf{w}_{k}},{\mathbf{m}_{t}}$ are jointly optimized. In the sub-problem 3, the radar receive beamforming vectors ${\mathbf{g}_{i}}$ and beamforming matrices ${\mathbf{w}_{k}},{\mathbf{m}_{t}}$ are given and IOS reflective and transmissive coefficient ${{\mathbf{\theta }}_{R}},{{\mathbf{\theta }}_{T}}$ are jointly optimized. Finally, the three sub-problems are optimized alternately until convergence is achieved.
\subsection{Sub-Problem 1: Radar Receive Vectors Design}
The optimal design for ${\mathbf{g}_{i}}$ should maximize the SINR of the echo signal of the $i$-th target. Without losing generality, the received beam energy is normalized here, and the sub-problem 1 can be expressed as
\begin{subequations}
\begin{align}
 \text{(P1)}:\quad &\mathop {\max }\limits_{{{\bf{g}}_i}} {\rm{  }}\frac{{\alpha _i^2{\bf{g}}_i^H{{\bf{S}}_i}{\bf{x}}{{\bf{x}}^H}{\bf{S}}_i^H{{\bf{g}}_i}}}{{{\bf{g}}_i^H\left( {\sum\limits_{o=1+\delta \left( {i - {O_R}} \right)\cdot{O_R},o\ne i}^{{{O}_{R}+\delta \left( {i - {O_R}} \right)\cdot{O_T}}} {\alpha _o^2{{\bf{S}}_o}{\bf{x}}{{\bf{x}}^H}{\bf{S}}_o^H}  + \sigma _i^2{\mathbf{I}_{{{N}_{r}}}}} \right){{\bf{g}}_i}}},\forall i,\\
\rm{s.t.}\quad &{{\bf{g}}_i}{\bf{g}}_i^H = 1,\forall i.
\end{align}
\end{subequations}
By defining ${{\mathbf{Q}}_{i}}= {{\mathbf{S}}_{i}}\mathbf{x}{{\mathbf{x}}^{H}}\mathbf{S}_{i}^{H}$ and
${{\mathbf{T}}_{i}}= \sum\limits_{o=1+\delta \left( {i - {O_R}} \right)\cdot{O_R},o\ne i}^{{{O}_{R}+\delta \left( {i - {O_R}} \right)\cdot{O_T}}}{\alpha _{o}^{2}{{\mathbf{S}}_{o}}\mathbf{x}{{\mathbf{x}}^{H}}\mathbf{S}_{o}^{H}}+\sigma _{i}^{2}{\mathbf{I}_{{{N}_{r}}}}$,
the optimal ${\mathbf{g}_{i}}$ can be expressed as
\begin{align}
\mathbf{g}_{i}^{*}=\underset{{{\mathbf{g}}_{i}}}{\mathop{\arg \max }}\,\frac{\mathbf{g}_{i}^{H}{{\mathbf{Q}}_{i}}{{\mathbf{g}}_{i}}}{\mathbf{g}_{i}^{H}{{\mathbf{T}}_{i}}{{\mathbf{g}}_{i}}},\forall i.
\end{align}
It’s obvious that ${{\mathbf{Q}}_{i}}$ is a Hermite matrix, and ${{\mathbf{T}}_{i}}$ is a positive definite Hermite matrix. Based on eigenvalue decomposition, ${{\mathbf{T}}_{i}}$ can be represented as
\begin{align}
{{\bf{T}}_i} = {{\bf{U}}_i}{\text{diag}}\left( {{\lambda _{i,1}},{\lambda _{i,2}}, \cdots {\lambda _{i,{N_r}}}} \right){\bf{U}}_i^H = {\left( {{{\bf{U}}_i}{\text{diag}}\left( {\sqrt {{\lambda _{i,1}}} ,\sqrt {{\lambda _{i,2}}} , \cdots \sqrt {{\lambda _{i,{N_r}}}} } \right){\bf{U}}_i^H} \right)^2},\forall i,
\end{align}
where ${{\mathbf{U}}_{i}}$ is a unitary matrix, ${{\lambda }_{i,1}},{{\lambda }_{i,2}},\cdots {{\lambda }_{i,{N_r}}}$ are ${N_r}$ positive eigenvalues of ${{\mathbf{U}}_{i}}$. By defining
${{\mathbf{B}}_{i}}=\left( {{\mathbf{U}}_{i}}\text{diag}\left( \sqrt{{{\lambda }_{i,1}}},\sqrt{{{\lambda }_{i,2}}},\cdots \sqrt{{{\lambda }_{{i,{N}_{r}}}}} \right)\mathbf{U}_{i}^{H} \right)$, which is also a definite Hermite matrix, its inverse matrix is still a Hermite matrix. Then, by defining ${{\mathbf{f}}_{i}}={{\mathbf{B}}_{i}}{{\mathbf{g}}_{i}}$, the problem of finding optimal ${{\mathbf{g}}_{i}}$ can be equivalent to finding optimal ${{\mathbf{f}}_{i}}$ and the problem (P1) can be represent as
\begin{subequations}
\begin{align}
 \text{(P1.1)}:\qquad &\underset{{{\mathbf{f}}_{i}}}{\mathop{\max }}\,\text{ }\mathbf{f}_{i}^{H}\mathbf{B}_{i}^{-1}{{\mathbf{Q}}_{i}}\mathbf{B}_{i}^{-1}{{\mathbf{f}}_{i}},\forall i,\\
\rm{s.t.}\qquad &\mathbf{f}_{i}^{H}{{\mathbf{f}}_{i}}={{r}_{i}},\forall i,
\end{align}
\end{subequations}
where ${{r}_{i}}$ is a constant whose value does not affect the solution of the problem. 
By constructing Lagrange functions 
$L\left( {{\mathbf{f}}_{i}},\lambda  \right)=\mathbf{f}_{i}^{H}\mathbf{B}_{i}^{-1}{{\mathbf{Q}}_{i}}\mathbf{B}_{i}^{-1}{{\mathbf{f}}_{i}}+\lambda \left( \mathbf{f}_{i}^{H}{{\mathbf{f}}_{i}}-{{r}_{i}} \right)$, the optimal ${{\mathbf{f}}_{i}}$ can be obtained by calculating the stagnation point method, i.e., ${\partial L\left( {{\mathbf{f}}_{i}},\lambda  \right)}/{\partial {{\mathbf{f}}_{i}}}\;=0\Rightarrow \mathbf{B}_{i}^{-1}{{\mathbf{Q}}_{i}}\mathbf{B}_{i}^{-1}{{\mathbf{f}}_{i}}=\lambda {{\mathbf{f}}_{i}}$,
which means the optimal ${{\mathbf{f}}_{i}}$ is the eigenvector corresponding to the maximum eigenvalue of ${\bf{B}}_i^{ - 1}{{\bf{Q}}_i}{\bf{B}}_i^{ - 1}$ and $\mathbf{g}_{i}^{*}=\mathbf{B}_{i}^{-1}\mathbf{f}_{i}^{*}$. Therefore the optimal minimum SINR of multi-target in the $m$-th after solving problem (P1) can be expressed as
\begin{align}\label{subop1}
q_{m,1}^{*}=\underset{i}{\mathop{\min }}\,\frac{{{\left( \mathbf{g}_{i}^{*} \right)}^{H}}{{\mathbf{Q}}_{i}}\mathbf{g}_{i}^{*}}{{{\left( \mathbf{g}_{i}^{*} \right)}^{H}}{{\mathbf{T}}_{i}}\mathbf{g}_{i}^{*}}.
\end{align}

\subsection{Sub-Problem 2: Beamforming Design of Transmitting Antenna}
In this subsection, given the radar receive beamforming vectors  ${{\mathbf{g}}_{i}}$ and IOS reflective and transmissive coefficient ${{\mathbf{\theta }}_{R}},{{\mathbf{\theta }}_{T}}$, beamforming matrices ${{\mathbf{W}}_{k}}$ and ${{\mathbf{m}}_{t}}$ are jointly optimized. then the problem (P0) can be written as the problem (P2), which is expressed as
\begin{subequations}\label{p2}
\begin{align}
  \text{(P2)}:\quad & \underset{\begin{smallmatrix} 
 {{\mathbf{W}}_{k}},{{\mathbf{m}}_{t}} \\ 
\end{smallmatrix}}{\mathop{\max }}\,\underset{i}{\mathop{\min}}\,\text{  SIN}{{\text{R}}_{i}}, \\ 
 \rm{s.t.}\quad &\text{    tr}\left( \sum\limits_{k=1}^{K}{{{\mathbf{W}}_{k}}\mathbf{W}_{k}^{H}}+\sum\limits_{t=1}^{{{N}_{T}}}{{{\mathbf{m}}_{t}}\mathbf{m}_{t}^{H}} \right)\le {{P}_{\max }}, \\ 
 &{{R}_{k}}\ge {{R}_{\text{th}}},\forall k. 
\end{align}
\end{subequations}
Considering that problem (P2) is an optimization problem with respect to ${{\mathbf{W}}_{k}},{{\mathbf{m}}_{t}}$ and ${{\mathbf{W}}_{k}},{{\mathbf{m}}_{t}}$ in the objective function appears in quadratic form, we use SDR to solve. Specifically,
by defining ${{\mathbf{\tilde{W}}}_{k}}={{\mathbf{W}}_{k}}\mathbf{W}_{k}^{H}$ and 
${{\mathbf{\tilde{M}}}_{t}}={{\mathbf{m}}_{t}}\mathbf{m}_{t}^{H}$, quadratic optimization problems can be transformed into linear problems. To ensure that the solved ${{\mathbf{W}}_{k}}$ and ${{\mathbf{m}}_{t}}$ can be decomposed into the product of the corresponding dimensional beamforming matrix, additional rank constraints need to be introduced, i.e., $\text{rank}\left( {{\mathbf{\tilde{W}}}_{k}} \right)\le {{D}_{k}}$ and $\text{rank}\left( {{\mathbf{\tilde{M}}}_{t}} \right)=1$. Because the objective function is in the max-min form and includes fractional programming, ${{q}_{m,x}}$ is introduced as an auxiliary variable, which represents the minimum SINR of multi-target when solving problem (P$x$) in the $m$-th iteration and problem (P2) can be represented as
\begin{subequations}\label{p2.1}
\begin{align}
 \text{(P2.1)}:\quad & \underset{{{{{\bf{\tilde W}}}_k}},{{{{\bf{\tilde M}}}_t}},{{q}_{m,2.1}}}{\mathop{\max }}\,{{q}_{m,2.1}}, \\ 
 \rm{s.t.}\quad & {\text{tr}}\left( \sum\limits_{k=1}^{K}{{{{\mathbf{\tilde{W}}}}_{k}}}+\sum\limits_{t=1}^{{{N}_{T}}}{{{{\mathbf{\tilde{M}}}}_{t}}} \right)\le {{P}_{\max }}, \\ 
 & \text{        }{{R}_{k}}\ge {{R}_{\text{th}}},\forall k, \\ 
 & \text{        tr}\left( \left( {{{\tilde{\mathbf{S}}}}_{i,i}}-{{q}_{m,2.1}}\sum\limits_{o=1+\delta \left( {i - {O_R}} \right)\cdot{O_R},o\ne i}^{{{O}_{R}+\delta \left( {i - {O_R}} \right)\cdot{O_T}}}{{{{\tilde{\mathbf{S}}}}_{i,o}}} \right)\left( \sum\limits_{k=1}^{K}{{{{\mathbf{\tilde{W}}}}_{k}}}+\sum\limits_{t=1}^{{{N}_{T}}}{{{{\mathbf{\tilde{M}}}}_{t}}} \right) \right)\ge {{q}_{m,2.1}}{{c}_{i}},\forall i, \\ 
 & \text{        rank}\left( {{{\mathbf{\tilde{W}}}}_{k}} \right)\le {{D}_{k}},\forall k, \\ 
 & \text{        rank}\left( {{{\mathbf{\tilde{M}}}}_{t}} \right)=1,\forall t,
\end{align}
\end{subequations}
where ${{\mathbf{\tilde{S}}}_{i,i}}=\alpha _{i}^{2}\mathbf{S}_{i}^{H}{\mathbf{g}_{i}}\mathbf{g}_{i}^{H}{{\mathbf{S}}_{i}},{{\mathbf{\tilde{\mathbf{S}}}}_{i,o}}=\alpha _{o}^{2}\mathbf{S}_{o}^{H}{\mathbf{g}_{i}}\mathbf{g}_{i}^{H}{{\mathbf{S}}_{o}},{{c}_{i}}=\sigma _{i}^{2}\text{tr}\left( {{\mathbf{g}}_{i}}\mathbf{g}_{i}^{H}{{\mathbf{I}}_{{{N}_{r}}}} \right)$. Because constraint (\ref{p2.1}d) is joint nonconvex with respect to ${{\mathbf{\tilde{W}}}_{k}},{{\mathbf{\tilde{M}}}_{t}},{{q}_{m,2.1}}$, we approximate ${{q}_{m,2.1}}$ on the left side of the inequality with a fixed point $q_{m,1}^{*}$, which can be calculated according to Eq. (\ref{subop1}) \cite{9163260}. The problem (P2.1) can be approximated as problem (P2.2) as follows
\begin{subequations}\label{p2.2}
\begin{align}
 \text{(P2.2)}:\quad & \underset{{{{\mathbf{\tilde{W}}}}_{k}},{{{\mathbf{\tilde{M}}}}_{t}},{{q}_{m,2.2}}}{\mathop{\max }}\,\text{  }{{q}_{m,2.2}}, \\ 
 \rm{s.t.}\quad &\text{         tr}\left( \left( {{{\mathbf{\tilde{S}}}}_{i,i}}-{{q}_{m,1}}\sum\limits_{o=1+\delta \left( {i - {O_R}} \right)\cdot{O_R},o\ne i}^{{{O}_{R}+\delta \left( {i - {O_R}} \right)\cdot{O_T}}}{{{{\mathbf{\tilde{S}}}}_{i,o}}} \right)\left( \sum\limits_{k=1}^{K}{{{{\mathbf{\tilde{W}}}}_{k}}}+\sum\limits_{t=1}^{{{N}_{T}}}{{{{\mathbf{\tilde{M}}}}_{t}}} \right) \right)\ge {{q}_{m,2.2}}{{c}_{i}},\forall i, \\ 
 & \text{(\ref{p2.1}b)}, \text{(\ref{p2.1}c)}, \text{(\ref{p2.1}e)}, \text{(\ref{p2.1}c)},
\end{align}
\end{subequations}
where constraint (\ref{p2.2}b) is joint convex with respect to ${{\mathbf{\tilde{W}}}_{k}},{{\mathbf{\tilde{M}}}_{t}}, {{q}_{m,2.2}}$.

$\textbf{Proposition 1:}$ After solving problem (P2.2) in the $m$-th iteration, $\mathbf{\tilde{W}}_{k}^{\left(m,2.2 \right)},\mathbf{\tilde{M}}_{t}^{\left(m,2.2 \right)}$ can be obtained, ${{q}_{n,2.1}}\left( \mathbf{\tilde{W}}_{k}^{\left(m,2.2 \right)},\mathbf{\tilde{M}}_{t}^{\left(m,2.2 \right)} \right)\ge {{q}_{m,1}}$ always holds which ensures the convergence of the algorithm.

\emph{Proof}: The proof is given in Appendix A. $\hfill\blacksquare$

Next, (\ref{p2}c) is nonconvex with respect to ${{\mathbf{\tilde{W}}}_{k}},{{\mathbf{\tilde{M}}}_{t}}$ and needs to be approximated. According to the properties of the determinant, the achievable rate of user $k$ can be expressed as
\begin{align}
\log \det \left( \sigma _{k}^{2}{{\mathbf{I}}_{{{M}_{k}}}}+\sum\limits_{k=1}^{K}{{{\mathbf{H}}_{k}}{{{\mathbf{\tilde{W}}}}_{k}}\mathbf{H}_{k}^{H}}+\sum\limits_{t=1}^{{{N}_{T}}}{{{\mathbf{H}}_{k}}{{{\mathbf{\tilde{M}}}}_{t}}\mathbf{H}_{k}^{H}} \right)-\log \det \left( {{\mathbf{J}}_{k}} \right),\forall k,
\end{align}
where the first term is already a concave function and the second term is a concave function. We can apply the first-order Taylor expansion to the convex function to get its lower bound. Specially, $\log \det \left( {{\mathbf{J}}_{k}} \right)$ can be approximated as follows
\begin{equation}
\begin{split}
& \log \det \left( {{{\bf{J}}_k}} \right) \le \log \det \left( {\mathop {{{\bf{J}}_k}}\limits^{\left( m \right)} } \right)\\
&+{\rm{tr}}\left( {\mathop {{\bf{J}}_k^{ - 1}}\limits^{\left( m \right)} \left( {\sum\limits_{l = 1,l \ne k}^K {{{\bf{H}}_k}{{{\bf{\tilde W}}}_l}{\bf{H}}_k^H}  + \sum\limits_{t = 1}^{{N_T}} {{{\bf{H}}_k}{{{\bf{\tilde M}}}_t}{\bf{H}}_k^H} } \right)} \right)\\
& - {\rm{tr}}\left( {\mathop {{\bf{J}}_k^{ - 1}}\limits^{\left( m \right)} \left( {\sum\limits_{l = 1,l \ne k}^K {{{\bf{H}}_k}{{\mathop {{\bf{\tilde W}}}\limits^{\left( m \right)} }_l}{\bf{H}}_k^H}  + \sum\limits_{t = 1}^{{N_T}} {{{\bf{H}}_k}{{\mathop {{\bf{\tilde M}}}\limits^{\left( m \right)} }_t}{\bf{H}}_k^H} } \right)} \right)\triangleq {{\left( \log \det \left( \overset{\left( m \right)}{\mathop{{{\mathbf{J}}_{k}}}}\, \right) \right)}^{ub}},\forall k,
\end{split}
\end{equation}
where $\left( \overset{\left( m \right)}{\mathop{{{\mathbf{J}}_{k}}}}\,,\overset{\left( m \right)}{\mathop{\mathbf{J}_{k}^{-1}}}\,,\overset{\left( m \right)}{\mathop{{{{\mathbf{\tilde{W}}}}_{k}}}}\,,\overset{\left( m \right)}{\mathop{{{{\mathbf{\tilde{M}}}}_{t}}}}\, \right)$ represents fixed points at the $m$-th iteration. Constraint (\ref{p2}c) can be represent as 
\begin{equation}\label{rkconvex}
\log \det \left( \sigma _{k}^{2}{{\mathbf{I}}_{{{M}_{k}}}}+\sum\limits_{k=1}^{K}{{{\mathbf{H}}_{k}}{{{\mathbf{\tilde{W}}}}_{k}}\mathbf{H}_{k}^{H}}+\sum\limits_{t=1}^{{{N}_{T}}}{{{\mathbf{H}}_{k}}{{{\mathbf{\tilde{M}}}}_{t}}\mathbf{H}_{k}^{H}} \right)-{{\left( \log \det \left( \overset{\left( m \right)}{\mathop{{{\mathbf{J}}_{k}}}}\, \right) \right)}^{ub}},\forall k.
\end{equation}
Constraint (\ref{rkconvex}) is a convex constraint. Next, because rank constraints (\ref{p2.1}e) and (\ref{p2.1}d) are nonconvex, we first use the following $\textbf{Proposition 2}$ \cite{sun2018rank} to handle the rank-$n$ constraint.

$\textbf{Proposition 2:}$ As for a matrix ${\bf{X}} \in S_ + ^n$, the eigenvalues of which are ${{\lambda }_{1}},{{\lambda }_{2}},\cdots ,{{\lambda }_{n}}$ and $0\le {{\lambda }_{1}}\le {{\lambda }_{2}}\le \cdots {{\lambda }_{n}}$. The $(r+1)$-th largest eigenvalue ${{\lambda }_{n-r}}$ of matrix ${\bf{X}}$ is no greater than if and only if $e{{\bf{I}}_{n-r}}-{{\bf{V}}^{H}}{\bf{X}}{\bf{V}}\ge 0$, where ${{\bf{I}}_{n-r}}$ is the identity matrix with dimension $n-r$ and ${\bf{V}}\in {{\mathbb{C}}^{n\times n-r}}$ are the eigenvectors corresponding to the $n-r$ smallest eigenvalues of ${\bf{X}}$.

\emph{Proof}: Assuming that the eigenvalues of ${\bf{X}}$ are sorted in descending order, $e{{\bf{I}}_{n-r}}-{{\bf{V}}^{H}}{\bf{X}}{\bf{V}}$ is a diagonal matrix whose diagonal elements are $\left[ e-{{\lambda }_{n-r}},e-{{\lambda }_{n-r-1}},\cdots e-{{\lambda }_{1}} \right]$. $e{{\bf{I}}_{n-r}}-{{\bf{V}}^{H}}{\bf{X}}{\bf{V}}$is a semi-positive definite matrix if and only if $0\le e-{{\lambda }_{n-r}}\le e-{{\lambda }_{n-r-1}}\le e-{{\lambda }_{1}}$, which indicates 
${{\lambda }_{n-r}}$ is no greater than $e$.

According to $\textbf{Proposition 2}$, when $e=0$, $\text{rank}\left( {\bf{X}} \right)\le r$ if and only if $e{{\bf{I}}_{n-r}}-{{\bf{V}}^{H}}{\bf{X}}{\bf{V}}\ge 0$. Therefore, constraint (\ref{p2.1}e) can be equivalently represented as ${{e}_{k}}{{\bf{I}}_{{{N}_{t}}-{{D}_{k}}}}-{{\bf{V}}_k^{H}}{{\tilde{\bf{W}}}_{k}}{\bf{V}}_k\ge 0,\forall k$. However, because it’s impossible to obtain eigenvector of ${{\tilde{\bf{W}}}_{k}}$ before solving problem (P2.2), we satisfy (\ref{p2.1}e) approximately through iterative rank minimization methods. Specifically, the ${\bf{V}}_k$ obtained in $m-1$-th iteration is used to approximate the in $m$-th iteration and the problem can be expressed as
\begin{subequations}\label{p2.3}
\begin{align}
 \text{(P2.3)}:\quad & \underset{{{{\tilde{\bf{W}}}}_{k}},{{{\tilde{\bf{M}}}}_{t}},{{q}_{m,2.2}},{{e}_{k}}}{\mathop{\max }}\,{{q}_{n,2.2}}-\sum\limits_{k=1}^{K}{{{\gamma }_{k}}{{\left( \gamma  \right)}^{m}}{{e}_{k}}},\\ 
 \rm{s.t.}\quad &{{e}_{k}}{{\bf{I}}_{{{N}_{t}}-{{D}_{k}}}}-\overset{\left( m \right)}{\mathop{{{\bf{V}}_k^{H}}}}\,{{{\tilde{\bf{W}}}}_{k}}\overset{\left( m \right)}{\mathop{{\bf{V}}_k}}\,\ge 0,\forall k,\\
 & {{e}_{k}}\le \overset{\left( m \right)}{\mathop{{{e}_{k}}}},\forall k,\\
 & \text{(\ref{p2.1}b)}, \text{(\ref{p2.1}f)}, \text{(\ref{p2.2}b)}, \text{(\ref{rkconvex})}, 
\end{align}
\end{subequations}
where ${{\gamma }_{0}}$ represents initial penalty factor, $\gamma \left( \gamma >1 \right)$ represents the exponential explosion factor, $\overset{\left( m \right)}{\mathop{\bf{V}}_k}\,$ represents is eigenvectors corresponding to the smallest ${{N}_{t}}-{{D}_{k}}$ eigenvalues of  given $\overset{\left( m \right)}{\mathop{{{{\tilde{\bf{W}}}}_{k}}}}\,$. Constraint (\ref{p2.3}c) ensures the monotonicity of ${{e}_{k}}$ after each iteration, in other words, minimize the value of ${{e}_{k}}$ as much as possible through penalty terms and constraints, so that ${{e}_{k}}$ gradually converges to 0. When ${{e}_{k}}$ approaches 0, that the $n-r+1$-th largest eigenvalue of a also approaches 0, making $\text{rank}\left( {{{\tilde{\bf{W}}}}_{k}} \right)\le {{D}_{k}}$ hold. Constraint (\ref{p2.2}c) is a rank-one constraint, which can be solved through the difference of convex algorithm \cite{9570143} and we omit the specific algorithm here. After completing the problem (P2.3) solution, ${{\mathbf{\tilde{W}}}_{k}}$ can be decomposed as
\begin{align}
{{\tilde{\bf{W}}}_{k}}={{\bf{U}}_{k}}{{{\bf{\Sigma}} }_{{{r}_{k}}}}{\bf{U}}_{k}^{H},
\end{align}
where ${{r}_{k}}$ is the rank of ${{\tilde{\bf{W}}}_{k}}$, ${{{\bf{\Sigma}} }_{{{r}_{k}}}}\in {{\mathbb{R}}^{{{r}_{k}}\times {{r}_{k}}}}$ is a diagonal matrix, the diagonal elements of which are ${{r}_{k}}$ positive eigenvalues of ${{\tilde{W}}_{k}}$ and ${{\bf{U}}_{k}}\in {{\mathbb{R}}^{{{N}_{t}}\times {{r}_{k}}}},{\bf{U}}_{k}^{H}{{\bf{U}}_{k}}={{\bf{I}}_{{{r}_{k}}}}$. If ${{r}_{k}}={{D}_{k}}$, the beamforming matrix for the multi-stream communication of user $k$ can be designed as ${{\bf{W}}_{k}}={\bf{U}}{\bf{\Sigma}} _{{{r}_{k}}}^{{1}/{2}\;}$. If ${{r}_{k}}<{{D}_{k}}$, select some column vector from ${\bf{U}}\Sigma _{{{r}_{k}}}^{{1}/{2}\;}$ for splitting in amplitude. The split column vector replaces the original column vector to expand the number of matrix columns. The total energy of split vectors should be equal to the energy of the original vector. For example, if ${{r}_{k}}+1={{D}_{k}}$, ${\bf{U}}{\bf{\Sigma}}_{{{r}_{k}}}^{{1}/{2}\;}=\left[ {{v}_{1}},{{v}_{2}},\cdots {{v}_{{{r}_{k}}}} \right]\in {{\mathbb{C}}^{{{N}_{t}}\times {{r}_{k}}}}$, a feasible design method for ${{\bf{W}}_{k}}$ is ${{\bf{W}}_{k}}=\left[ {{v}_{1}},{{v}_{2}},\cdots {{{v}_{{{r}_{k}}}}}/{\sqrt{2}}\;,{{{v}_{{{r}_{k}}}}}/{\sqrt{2}}\; \right]\in {{\mathbb{C}}^{{{N}_{t}}\times {{D}_{k}}}}$. 
\subsection{Sub-Problem 3: Design of Reflective and Transmissive Coefficients for IOS}
In this subsection, given the radar receive beamforming vectors ${\mathbf{g}_{i}}$ and, beamforming matrices
${\mathbf{W}_{k}},{\mathbf{m}_{t}}$, ${\mathbf{\theta}_{R}},{\mathbf{\theta}_{T}}$ are jointly optimized. then the problem (P0) can be written as the problem (P3), which is expressed as
\begin{subequations}\label{p3}
\begin{align}
 \text{(P3)}:\quad & \underset{{{{\bf{\Theta}} }_{r}},{{{\bf{\Theta}} }_{t}}}{\mathop{\max }}\,\underset{i}{\mathop{\min}} \quad \text{SIN}{{\text{R}}_{i}}, \\ 
\rm{s.t.}\quad & {{R}_{k}}\ge {{R}_{\text{th}}},\forall k ,\\ 
 & \alpha _{n}^{t},\alpha _{n}^{r}\in \left[ 0,1 \right],\alpha _{n}^{t}+\alpha _{n}^{r}=1,\forall n, \\ 
 & \beta _{n}^{t},\beta _{n}^{r}\in \left[ 0,2\pi  \right),\forall n.
\end{align}
\end{subequations}
The problem (P3) is challenging for the following reasons. First, constraint (\ref{p3}b) is non-convex with respect to ${\mathbf{\theta}_{T}},{\mathbf{\theta}_{R}}$. Then, the traditional IOS coefficient optimization problem based on MISO scenario, where IOS coefficient matrix can be expressed as vector through the product exchange of vector and diagonal matrix. However, in MIMO communication, the expressions for reachable rates are all matrix products, which makes it difficult to apply the SDR algorithm to process quadratic forms of IOS coefficient directly. Therefore, we use the weighted MMSE approach \cite{5756489} to represent constraint (\ref{p3}b).

$\textbf{Proposition 3:}$ The constraint (\ref{p3}b) to  ensure the achievable rate of each user is equivalent to the weighted MSE constraint when the weight matrix satisfies ${\mathbf{V}_{k}}=\mathbf{E}_{k}^{-1}$, which can be expressed as
\begin{align}
{{f}_{k}}=\text{tr}\left( {\mathbf{V}_{k}}{\mathbf{E}_{k}} \right)-\log \det \left( {\mathbf{V}_{k}} \right)\le {{D}_{k}}-{{R}_\text{th}},\forall k,
\end{align}
with
\begin{equation}
\begin{split}
& {\mathbf{E}_{k}}=E\left[ \left( {\mathbf{d}_{k}}-{{{\hat{\mathbf{d}}}}_{k}} \right){{\left( {\mathbf{d}_{k}}-{{{\hat{\mathbf{d}}}}_{k}} \right)}^{H}} \right] \\ 
& =E\left[\left( \left( {\mathbf{d}_{k}}-{\mathbf{G}_{k}}{\mathbf{H}_{k}}{\mathbf{W}_{k}}{\mathbf{d}_{k}} \right)-{\mathbf{G}_{k}}{\mathbf{H}_{k}}\left( \sum\limits_{i=1,i\ne k}^{K}{{\mathbf{W}_{i}}{\mathbf{d}_{i}}}+\sum\limits_{t=1}^{{{N}_{T}}}{{\mathbf{m}_{t}}{{s}_{t}}} \right)+{\mathbf{G}_{k}}{\mathbf{n}_{k}} \right) \right.\\
& \left.{{\left( \left( {\mathbf{d}_{k}}-{\mathbf{G}_{k}}{\mathbf{H}_{k}}{\mathbf{W}_{k}}{\mathbf{W}_{k}} \right)-{\mathbf{G}_{k}}{\mathbf{H}_{k}}\left( \sum\limits_{i=1,i\ne k}^{K}{{\mathbf{W}_{i}}{\mathbf{d}_{i}}}+\sum\limits_{t=1}^{{{N}_{T}}}{{\mathbf{m}_{t}}{{s}_{t}}} \right)+{\mathbf{G}_{k}}{\mathbf{n}_{k}} \right)}^{H}} \right], \forall k,
\end{split}
\end{equation}
where ${\mathbf{V}_{k}}$ represents the MSE weight matrix and ${\mathbf{E}_{k}}$ is the MSE matrix of user $k$. ${\mathbf{G}_{k}}$ represents the MIMO communication receiving beamforming matrix.

\emph{Proof}: The proof is given in Appendix B. $\hfill\blacksquare$

However, The entire optimization problem introduces two additional variables ${\mathbf{G}_{k}}$ and ${\mathbf{V}_{k}}$ through weighted MMSE approach. Therefore, we use alternating optimization methods to update these three variables. The optimal $\mathbf{G}_{k}^{*}$ and $\mathbf{V}_{k}^{*}$ can be obtained by calculating $\partial {{{f}_{k}}}/{\partial }\mathbf{G}_{k}^{*}=0$ and $\partial {{{f}_{k}}}/{\partial }\mathbf{V}_{k}^{*}=0$ respectively. i.e.,
\begin{align}\label{gkvk}
\mathbf{G}_{k}^{*}=\frac{\mathbf{W}_{k}^{H}\mathbf{H}_{k}^{H}}{\sum\limits_{i=1}^{K}{{\mathbf{H}_{k}}{\mathbf{W}_{i}}\mathbf{W}_{i}^{H}\mathbf{H}_{k}^{H}}+\sum\limits_{t=1}^{{{N}_{T}}}{{\mathbf{H}_{k}}{\mathbf{m}_{t}}\mathbf{m}_{t}^{H}\mathbf{H}_{k}^{H}}+\sigma _{k}^{2}\mathbf{I}_{{M}_{k}}},\mathbf{V}_{k}^{*}= \mathbf{E}_{k}^{-1}.
\end{align}
Then, we use SDR to handle the quadratic forms of ${{\theta }_{T}}$ and ${{\theta }_{R}}$. Utilizing the properties of trace, $\text{tr}\left( {\mathbf{V}_{k}}{\mathbf{E}_{k}} \right)$ can be expressed as 
\begin{equation}
\begin{split}
& \text{tr}\left( {{\mathbf{V}}_{k}}{{\mathbf{E}}_{k}} \right)=\text{tr}\left( {{\mathbf{V}}_{k}}+E\left[ \sum\limits_{i=1}^{K}{\mathbf{d}_{i}^{H}\mathbf{W}_{i}^{H}\mathbf{H}_{k}^{H}\mathbf{G}_{k}^{H}{{\mathbf{V}}_{k}}{{\mathbf{G}}_{k}}{{\mathbf{H}}_{k}}{{\mathbf{W}}_{i}}{{\mathbf{d}}_{i}}}-\mathbf{d}_{k}^{H}\mathbf{W}_{k}^{H}\mathbf{H}_{k}^{H}\mathbf{G}_{k}^{H}{{\mathbf{V}}_{k}}{{\mathbf{d}}_{k}} \right. \right) \\ 
& \left. -\left. \mathbf{d}_{k}^{H}{{\mathbf{V}}_{k}}{{\mathbf{G}}_{k}}{{\mathbf{H}}_{k}}{{\mathbf{W}}_{k}}{{\mathbf{d}}_{k}}+\sum\limits_{t=1}^{{{N}_{T}}}{s_{t}^{H}\mathbf{m}_{t}^{H}\mathbf{H}_{k}^{H}\mathbf{G}_{k}^{H}{{\mathbf{V}}_{k}}{{\mathbf{G}}_{k}}{{\mathbf{H}}_{k}}{\mathbf{m}_{t}}{{s}_{t}}} \right]+\sigma _{k}^{2}{{\mathbf{V}}_{k}}{{\mathbf{G}}_{k}}\mathbf{G}_{k}^{H} \right).
\end{split}
\end{equation}
By applying the change of variables $\mathbf{Z}_{k,k}^{H}=\text{diag}\left( \tilde{\mathbf{d}}_{k}^{H} \right)\mathbf{H}_{k,\text{UR}}^{H}\mathbf{G}_{k}^{H}$, $\mathbf{Z}_{k,i}^{H}=\text{diag}\left( \tilde{\mathbf{d}}_{i}^{H} \right)\mathbf{H}_{k,\text{UR}}^{H}\mathbf{G}_{k}^{H}$, $\tilde{\mathbf{s}}_{t}^{H}=s_{t}^{H}\mathbf{m}_{t}^{H}\mathbf{H}_{\text{RB}}^{H}$ and $\tilde{\mathbf{d}}_{k}^{H}\mathbf{\Theta} _{{S_K}\left( k \right)}^{H}=\mathbf{\theta}_{{S_K}\left( k \right)}^{H}\text{diag}\left( \tilde{\mathbf{d}}_{k}^{H} \right)$, where $\tilde{\mathbf{d}}_{k}^{H}=\mathbf{d}_{k}^{H}\mathbf{W}_{k}^{H}\mathbf{H}_{\text{RB}}^{H}$ and  $\tilde{\mathbf{s}}_{t}^{H}=s_{t}^{H}\mathbf{m}_{t}^{H}\mathbf{H}_{\text{RB}}^{H}$. Thus, ${{f}_{k}}$ is equivalent to
\begin{equation}\label{trve}
\begin{aligned}
& \text{tr}\left( {\mathbf{V}_{k}}+\sigma _{k}^{2}{\mathbf{V}_{k}}{\mathbf{G}_{k}}\mathbf{G}_{k}^{H} \right)-\log \det \left( {\mathbf{V}_{k}} \right) +\text{tr}\left( \sum\limits_{i=1}^{K}{\mathbf{\theta} _{{S_K}\left( k \right)}^{H}\mathbf{Z}_{k,i}^{H}{\mathbf{V}_{k}}{\mathbf{Z}_{k,i}}{{\mathbf{\theta} }_{{S_K}\left( k \right)}}} \right.\\
 &+\left. \sum\limits_{t=1}^{{{N}_{T}}}{\theta _{{S_K}\left( k \right)}^{H}\mathbf{Z}_{k,t}^{H}{\mathbf{V}_{k}}{\mathbf{Z}_{k,t}}}{\mathbf{\theta}_{{S_K}\left( k \right)}}-\mathbf{\theta} _{{S_K}\left( k \right)}^{H}\mathbf{Z}_{k,k}^{H}{\mathbf{V}_{k}}{\mathbf{d}_{k}}-\mathbf{d}_{k}^{H}{\mathbf{V}_{k}}{\mathbf{Z}_{k,k}}{\mathbf{\theta}_{{S_K}\left( k \right)}} \right).
\end{aligned}
\end{equation}
Eq. (\ref{trve}) is a non-convex quadratic function, which can be reformulated as a homogeneous quadratic function by introducing an auxiliary variable $t$. Specifically, Eq. (\ref{trve}) can be represented as $\text{tr}\left( {\mathbf{V}_{k}}+\sigma _{k}^{2}{\mathbf{v}_{k}}{\mathbf{G}_{k}}\mathbf{G}_{k}^{H}+E\left[ {\mathbf{A}_{1,k}} \right]{{{\tilde{\mathbf{\Theta} }}}_{k}} \right)$, where 
\begin{align}
{{\tilde{\mathbf{\Theta} }}_{k}}={{\tilde{\mathbf{\theta} }}_{k}}\tilde{\mathbf{\theta} }_{k}^{H},{{\tilde{\mathbf{\theta}}}_{k}}=\left[ {{\mathbf{\theta} }_{{S_K}\left( k \right)}},t \right],{\mathbf{A}_{1,k}}=\left[ \begin{matrix}
   \sum\limits_{i=1}^{K}{\mathbf{Z}_{k,i}^{H}{\mathbf{V}_{k}}{\mathbf{Z}_{k,i}}}+\sum\limits_{t=1}^{{{N}_{T}}}{\mathbf{Z}_{k,t}^{H}{\mathbf{V}_{k}}{\mathbf{Z}_{k,t}}} & -\mathbf{Z}_{k,k}^{H}{\mathbf{V}_{k}}{\mathbf{d}_{k}}  \\
   -\mathbf{d}_{k}^{H}{\mathbf{V}_{k}}{\mathbf{Z}_{k,k}} & 0
\end{matrix} \right].
\end{align}
In addition, the amplitude of the linear term needs to remain constant i.e., $\left| t \right|=1$.
Similarly, we use SDR to express the performance of radar system. By defining 
\begin{align}
& \mathbf{\tilde{g}}_{i,i}^{H}=\mathbf{g}_{i}^{H}\mathbf{A}\left( {{\theta }_{i,\text{OR,h}}},{{\theta }_{i,\text{OR,v}}},{{\theta }_{i,\text{RO,h}}} \right),\mathbf{\tilde{g}}_{i,o}^{H}=\mathbf{g}_{i}^{H}\mathbf{A}\left( {{\theta }_{o,\text{OR,h}}},{{\theta }_{o,\text{OR,v}}},{{\theta }_{o,\text{RO,h}}} \right), \\ 
 & {{\mathbf{Y}}_{i,i}}=\alpha _{i}^{2}\text{diag}\left( \mathbf{\tilde{g}}_{i,i}^{H} \right){{\mathbf{H}}_{RB}}\left( \sum\limits_{k=1}^{K}{{{{\mathbf{\tilde{W}}}}_{k}}}+\sum\limits_{t=1}^{{{N}_{T}}}{{{{\mathbf{\tilde{M}}}}_{t}}} \right)\mathbf{H}_{RB}^{H}\text{diag}\left( {{{\mathbf{\tilde{g}}}}_{i,i}} \right), \\ 
 & {{\mathbf{Y}}_{i,o}}=\sum\limits_{o=1,o\ne i}^{{{O}_{r}}}{\alpha _{o}^{2}\text{diag}\left( \mathbf{\tilde{g}}_{i,o}^{H} \right){{\mathbf{H}}_{RB}}\left( \sum\limits_{k=1}^{K}{{{{\mathbf{\tilde{W}}}}_{k}}}+\sum\limits_{t=1}^{{{N}_{T}}}{{{{\mathbf{\tilde{M}}}}_{t}}} \right)}\mathbf{H}_{RB}^{H}\text{diag}\left( {{{\mathbf{\tilde{g}}}}_{i,o}} \right),
\end{align}
and leveraging $\tilde{\mathbf{g}}_{i,i}^{H}\mathbf{\Theta }_{{{S}_{O}}\left( i \right)}^{H}=\mathbf{\theta }_{{{S}_{O}}\left( i \right)}^{H}\text{diag}\left( \mathbf{g}_{i,i}^{H} \right)$, $\tilde{\mathbf{g}}_{i,o}^{H}\mathbf{\Theta }_{{{S}_{O}}\left( i \right)}^{H}=\mathbf{\theta }_{{{S}_{O}}\left( i \right)}^{H}\text{diag}\left( \mathbf{g}_{i,o}^{H} \right)$, the SINR of the $i$-th target echo signal can be expressed as
\begin{align}
\text{SIN}{{\text{R}}_{i}}=\frac{\text{tr}\left( {\mathbf{Y}_{i,i}}{{\mathbf{\theta }}_{{{S}_{O}}\left( i \right)}}\mathbf{\theta }_{{{S}_{O}}\left( i \right)}^{H} \right)}{\text{tr}\left( {\mathbf{Y}_{i,o}}{{\mathbf{\theta }}_{{{S}_{O}}\left( i \right)}}\mathbf{\theta }_{{{S}_{O}}\left( i \right)}^{H} \right)},\forall i.
\end{align}
To ensure that the dimensions of the optimization variables are the same as those in communication constraints, we define ${{\tilde{\mathbf{\Theta} }}_{i}}$, ${{\tilde{\mathbf{\theta}}}_{i}}$, ${\mathbf{A}_{2,i}}$ and ${\mathbf{A}_{2,i}}$, which are
\begin{align}
{{\tilde{\mathbf{\Theta} }}_{i}}={{\tilde{\mathbf{\theta} }}_{i}}\tilde{\mathbf{\theta} }_{i}^{H},{{\tilde{\mathbf{\theta}}}_{i}}=\left[ {{\mathbf{\theta} }_{{S_O}\left( i \right)}},t \right],
{\mathbf{A}_{2,i}}=\left[ \begin{matrix}
   {\mathbf{Y}_{i,i}} & 0  \\
   0 & 0  \\
\end{matrix} \right],{\mathbf{A}_{3,i}}=\left[ \begin{matrix}
   {\mathbf{Y}_{i,o}} & 0  \\
   0 & 0  \\
\end{matrix} \right],\forall i.
\end{align}
Therefore, $\text{SIN}{{\text{R}}_{i}}$ can be represented as ${\text{tr}\left( {\mathbf{A}_{2,i}}{{\tilde{\mathbf{\Theta} }}_{i}} \right)}/{\text{tr}\left( {\mathbf{A}_{3,i}}{{\tilde{\mathbf{\Theta} }}_{i}} \right)}$. Finally, we use the same approximation method as sub-problem 2 to handle max-min problem in objective function and problem (P3) can be expressed as 
\begin{subequations}\label{p3.1}
\begin{align}
  \text{(P3.1)}:\qquad & \underset{{{{\tilde{\mathbf{\Theta} }}}_{i}},{{{\tilde{\mathbf{\Theta} }}}_{k}}, {{q}_{m,3.1}}}{\mathop{\max }}\,{{q}_{m,3.1}}, \\
 \rm{s.t.}\qquad &\text{tr}\left( {\mathbf{V}_{k}}+\sigma _{k}^{2}{\mathbf{V}_{k}}{\mathbf{G}_{k}}\mathbf{G}_{k}^{H}+E\left[ {\mathbf{A}_{1,k}} \right]{{{\tilde{\mathbf{\Theta} }}}_k} \right)-\log \det \left( {\mathbf{V}_{k}} \right)\le {{D}_{k}}-{{R}_\text{th}},\forall k, \\ 
 & \text{tr}\left( \left( {\mathbf{A}_{2,i}}-{{q}_{m,2.1}}{\mathbf{A}_{3,i}} \right)\tilde{\Theta }_i \right)\ge {{q}_{m,3.1}}{{c}_{i}},\forall i, \\ 
 & {{{\tilde{\mathbf{\Theta}}}}_{k}}\left[ N+1,N+1 \right]={{{\tilde{\mathbf{\Theta} }}}_{i}}\left[ N+1,N+1 \right]=1,\\
 & \text{(\ref{p3}c)}, \text{(\ref{p3}d)},
\end{align}
\end{subequations}
where ${{q}_{m,2.1}}$ denotes the minimum SINR which has been obtained after solving problem (P2.2) in the $m$-th iteration. ${{\tilde{\mathbf{\Theta}}}_{k}}\left[ N+1,N+1 \right]$ and ${{\tilde{\mathbf{\Theta}}}_{i}}\left[ N+1,N+1 \right]$ represent the element in $N+1$-th row and $N+1$-th column of ${{\tilde{\mathbf{\Theta}}}_{k}}$ and ${{\tilde{\mathbf{\Theta}}}_{i}}$ respectively. ${{q}_{m,1}}\le {{q}_{m,2.1}}\le {{q}_{m,3.1}}\le {{q}_{m+1,1}}$ always hold, which ensures the convergence of the entire algorithm and will be proved in Appendix A. After the above equivalence transformation and approximation, problem (P3.1) is a standard SDP problem and can be solved by using CVX toolbox.
\subsection{The Overall Joint Active And Passive Beamforming Design For IOS-Assisted ISAC Network}
In this subsection, we propose the overall joint radar receive vectors, transmit beamforming matrices, and IOS reflective and transmissive coefficients optimization algorithm and summarize it in \textbf{Algorithm 1}. We can respectively solve the problem (P1.1), (P2.3) and (P3.1) to obtain radar receive vectors, transmit beamforming matrices, and IOS reflective and transmissive coefficients. Finally, the three sub-problems are optimized alternately until the entire problem converges.
\begin{algorithm}[t]
	\caption{Joint Active And Passive Beamforming Design For IOS-Assisted ISAC Network} 
	\begin{algorithmic}[1]
		\State$\textbf{Input}$: Initialize feasible points $\overset{\left(1 \right)}{\mathop{{{\mathbf{g}}_{i}}}},\overset{\left( 1 \right)}{\mathop{{{{\mathbf{\tilde{W}}}}_{k}}}},\overset{\left( 1 \right)}{\mathop{{{{\mathbf{\tilde{M}}}}_{t}}}},\overset{\left( 1 \right)}{\mathop{{{\mathbf{\Theta }}_{R}}}},\overset{\left( 1 \right)}{\mathop{{{\mathbf{\Theta }}_{T}}}}$, the penalty
factor ${{\gamma }_{0}}$ and exponential explosion factor ${\gamma }$.
		\Repeat: outer loop
        \State Set iteration index $m=1$ for inner loop.
		\Repeat: inner loop
        \State Solve the problem (P1.1) to obtain radar receive vectors ${{\mathbf{g}}_{i}}$ and update $\overset{\left( m+1 \right)}{\mathop{{{\mathbf{g}}_{i}}}}$.
        \State Solve the problem (P2.3) to obtain transmit beamforming matrices ${{\mathbf{\tilde{W}}}_{k}},{{\mathbf{\tilde{M}}}_{t}}$ and updata $\overset{\left( m+1 \right)}{\mathop{{{{\mathbf{\tilde{W}}}}_{k}}}},\overset{\left( m+1 \right)} {\mathop{{{{\mathbf{\tilde{M}}}}_{t}}}}$.
        \State Solve the problem (P3.1) to obtain IOS reflective and transmissive coefficients ${{\mathbf{\Theta }}_{R}},{{\mathbf{\Theta }}_{T}}$ and updata $\overset{\left( m+1 \right)}{\mathop{{{\mathbf{\Theta }}_{R}}}},\overset{\left( m+1 \right)}{\mathop{{{\mathbf{\Theta }}_{T}}}},m=m+1$.
		\Until The fractional decrease of the objective function value is below a predefined threshold.
        \State Updata ${{\gamma }_{0}}={{\gamma }_{0}}\cdot \gamma$ and feasible points $\overset{\left(1 \right)}{\mathop{{{\mathbf{g}}_{i}}}},\overset{\left( 1 \right)}{\mathop{{{{\mathbf{\tilde{W}}}}_{k}}}},\overset{\left( 1 \right)}{\mathop{{{{\mathbf{\tilde{M}}}}_{t}}}},\overset{\left( 1 \right)}{\mathop{{{\mathbf{\Theta }}_{R}}}},\overset{\left( 1 \right)}{\mathop{{{\mathbf{\Theta }}_{T}}}}$ with current solutions. 
        \Until The constraint violation is below a predefined threshold.
        \State$\textbf{Output}$: $\mathbf{g}_{i}^{*},\mathbf{\tilde{W}}_{k}^{*},\mathbf{\tilde{M}}_{t}^{*},\mathbf{\Theta }_{R}^{*},\mathbf{\Theta }_{T}^{*}$.
	\end{algorithmic}
\end{algorithm}
\subsection{Proposed Solution for MS}
Compared with the ES protocol, the algorithm design of MS protocol only changes the design way of the IOS coefficient without changing that of radar receive beamforming and transit beamforming. Specifically, the amplitude of the reflective and transmissive coefficients of each element should be 0 or 1, which is a discrete binary constraint and can be represented as
\begin{align}\label{ms}
\alpha _{n,R}^{2}-{{\alpha }_{n,R}}=0,\alpha _{n,T}^{2}-{{\alpha }_{n,T}}=0.
\end{align}
The coefficient amplitude obtained by solving problem (P2) is between 0 and 1, leading the value of $\alpha _{n,R}^{2}-{{\alpha }_{n,R}}$ and $\alpha _{n,T}^{2}-{{\alpha }_{n,T}}$ is always less than 0. To satisfy Eq. (\ref{ms}), a penalty term can be introduced to make the coefficient amplitude approaches 0 or 1. Since the penalty term is nonconvex, it needs to be converted into convex function through first-order Taylor expansion, i.e.,
\begin{align}
& \alpha _{n,R}^{2}-{{\alpha }_{n,R}}\ge \left( \overset{\left( m \right)}{\mathop{2{{\alpha }_{n,R}}}}\,-1 \right){{\alpha }_{n,R}}-\overset{\left( m \right)}{\mathop{\alpha _{n,R}^{2}}}\,\triangleq \overset{\left( m \right)}{\mathop{{{\gamma }_{n,R}}}},\\ 
& \alpha _{n,T}^{2}-{{\alpha }_{n,T}}\ge \left( \overset{\left( m \right)}{\mathop{2{{\alpha }_{n,T}}}}\,-1 \right){{\alpha }_{n,T}}-\overset{\left( m \right)}{\mathop{\alpha _{n,T}^{2}}}\,\triangleq \overset{\left( m \right)}{\mathop{{{\gamma }_{n,T}}}}.  
\end{align}
where $\overset{\left( m \right)}{\mathop{{{\alpha }_{n,R}}}}$ and $\overset{\left( m \right)}{\mathop{{{\alpha }_{n,T}}}}$ are the given point in the $m$-th iteration. Therefore, the optimization problem can be represented as
\begin{align}
 \text{(P4)}: & \underset{{{\tilde{\mathbf{\Theta }}}_{k}},{{\tilde{\mathbf{\Theta }}}_{i}},{{q}_{m,3.1}}}{\mathop{\max }}\,{{q}_{m,3.1}}-{{\nu }_{0}}{{\left( \nu  \right)}^{m}}\sum\limits_{n=1}^{N}{\left( \overset{\left( m \right)}{\mathop{{{\gamma }_{n,R}}}}\,+\overset{\left( m \right)}{\mathop{{{\gamma }_{n,T}}}}\, \right)},\\ 
\rm{s.t.}&\quad\text{(\ref{p3}c)}, \text{(\ref{p3}d)}, \text{(\ref{p3.1}b)},\text{(\ref{p3.1}c)},\text{(\ref{p3.1}d)},
\end{align}
where ${{\nu }_{0}}$ represents the initial penalty factor and $\nu \left(\nu>1 \right)$ represents the exponential explosion factor. Then, problem (P4) is a standard convex optimization problem that can be solved by using CVX toolbox.
\subsection{Proposed Solution for TS}
When applying the TS protocol to the IOS system, the dual-function base stations serve communication users and sensed targets in reflective and transmissive areas periodically and The switching period is the same as that of IOS. The communications volume is linearly affected by transmission time, i.e., the achievable rate of $k$-th user will reduce to ${{\lambda }_{{{S}_{K}}\left( k \right)}}$  equivalently and can be expressed as
\begin{align}\label{rts}
R_{k}^{\text{TS}}={{\lambda }_{{{S}_{K}}\left( k \right)}}\log \det \left({{\mathbf{I}}_{{{D}_{k}}}}+\mathbf{W}_{k}^{H}\mathbf{H}_{k}^{H}\mathbf{J}_{k}^{-1}{{\mathbf{H}}_{k}}{{\mathbf{W}}_{k}} \right),\forall k.
\end{align}
Similarly, the signal energy reflected by the perceived target is linearly related to the accumulation time of the pulse, so the SINR of $i$-th target in TS protocol can be represented as
\begin{align}\label{sinrts}
\text{SINR}_{i}^{\text{TS}}=\frac{{{\lambda }_{{{S}_{O}}\left( i \right)}}\alpha _{i}^{2}{{\left| \mathbf{g}_{i}^{H}{{\mathbf{S}}_{i}}\mathbf{x} \right|}^{2}}}{\mathbf{g}_{i}^{H}\left( \sum\limits_{o=1+\delta \left( {i - {O_R}} \right)\cdot{O_R},o\ne i}^{{{O}_{R}+\delta \left( {i - {O_R}} \right)\cdot{O_T}}}{\alpha _{o}^{2}{{\mathbf{S}}_{o}}\mathbf{x}{{\mathbf{x}}^{H}}\mathbf{S}_{o}^{H}}+\sigma _{i}^{2}{{\mathbf{I}}_{{{N}_{r}}}} \right){{\mathbf{g}}_{i}}},\forall i,
\end{align}
where ${{\lambda }_{R}}$ and ${{\lambda }_{T}}$ are newly introduced optimization variables compared with problem (P0). Therefore, we still divide all optimization variables into three blocks which are $\left\{ {\mathbf{g}_{i}} \right\}$, $\left\{ {\mathbf{W}_{k}},{\mathbf{m}_{t}} \right\}$,    and $\left\{ {{\mathbf{\theta} }_{R}},{{\mathbf{\theta} }_{T}},{{\lambda }_{R}},{{\lambda }_{T}} \right\}$ to optimizing alternately. The algorithm of optimize $\left\{ {\mathbf{g}_{i}} \right\}$ and $\left\{ {\mathbf{W}_{k}},{\mathbf{m}_{t}} \right\}$ are the same as that of MS protocol and we consider optimizing $\left\{ {\mathbf{g}_{i}} \right\}$ and $\left\{ {\mathbf{W}_{k}},{\mathbf{m}_{t}} \right\}$. Recalling Eq. (\ref{rts}). and Eq. (\ref{sinrts}), the sub-problem can be expressed as
\begin{subequations}\label{p5}
\begin{align}
 \text{(P5)}:\quad & \underset{{{\tilde{\mathbf{\Theta}}}_{k}},{{\tilde{\mathbf{\Theta}}}_{i}}, {{q}_{m,3.1}},{{\lambda }_{r}},{{\lambda }_{t}}}{\mathop{\max }}\,{{q}_{m,3.1}}, \\
\rm{s.t.}\quad &\text{tr}\left( {\mathbf{V}_{k}}+\sigma _{k}^{2}{\mathbf{V}_{k}}{\mathbf{G}_{k}}\mathbf{G}_{k}^{H}+E\left[ {{A}_{1,k}} \right]{{\tilde{\mathbf{\Theta}}}_{k}} \right)-{{\lambda }_{{{S}_{K}}\left( k \right)}}\log \det \left( {\mathbf{V}_{k}} \right)\le {{D}_{k}}-{{R}_\text{th}},\forall k,\\ 
 & {{\lambda }_{{{S}_{O}}\left( i \right)}}\text{tr}\left( \left( {\mathbf{A}_{2,i}}-{{q}_{m,2.1}}{\mathbf{A}_{3,i}} \right){{\tilde{\mathbf{\Theta}}}_{i}} \right)\ge {{q}_{m,3.1}}{{c}_{i}},\forall i,\\ 
 & {{\lambda }_{r}}+{{\lambda }_{t}}=1,0\le {{\lambda }_{t}}\le 1,0\le {{\lambda }_{r}}\le 1,\\
 & \text{(\ref{p3}c)}, \text{(\ref{p3}d)}, \text{(\ref{p3.1}d)},
\end{align}
\end{subequations}
where (\ref{p5}b) is already a convex and (\ref{p5}c) is a bilinear constraint with respect to ${{\lambda }_{{{S}_{O}}\left( i \right)}}$ and ${{\tilde{\mathbf{\Theta}}}_{i}}$ and is a non-convex constraint. Therefore, we transform the left-hand side of (\ref{p5}c) into the following difference of convex (DC) functions and find its lower bound by first-order Taylor expansion. Specifically, ${{\lambda }_{{{S}_{O}}\left( i \right)}}\text{tr}\left( \left( {{\mathbf{A}}_{2,i}}-{{q}_{m,2.1}}{{\mathbf{A}}_{3,i}} \right){{{\mathbf{\tilde{\Theta }}}}_{i}} \right)$ can be expressed as
\begin{equation}
\begin{split}
&\frac{{{\left( {{\lambda }_{{{S}_{O}}\left( i \right)}}+\text{tr}\left( \left( {{\mathbf{A}}_{2,i}}-{{q}_{m,2.1}}{{\mathbf{A}}_{3,i}} \right){{{\mathbf{\tilde{\Theta }}}}_{i}} \right) \right)}^{2}}}{4}-\frac{{{\left( {{\lambda }_{{{S}_{O}}\left( i \right)}}-\text{tr}\left( \left( {{\mathbf{A}}_{2,i}}-{{q}_{m,2.1}}{{\mathbf{A}}_{3,i}} \right){{{\mathbf{\tilde{\Theta }}}}_{i}} \right) \right)}^{2}}}{4}\\ 
& \ge \frac{{{\left( \overset{\left( m \right)}{\mathop{{{\lambda }_{{{S}_{O}}\left( i \right)}}}}\,+\text{tr}\left( \left( {{\mathbf{A}}_{2,i}}-{{q}_{m,2.1}}{{\mathbf{A}}_{3,i}} \right)\overset{\left( m \right)}{\mathop{{{{\mathbf{\tilde{\Theta }}}}_{i}}}}\, \right) \right)}^{2}}}{4}
-\frac{{{\left( \overset{\left( m \right)}{\mathop{{{\lambda }_{{{S}_{O}}\left( i \right)}}}}\,-\text{tr}\left( \left( {{\mathbf{A}}_{2,i}}-{{q}_{m,2.1}}{{\mathbf{A}}_{3,i}} \right){{{\mathbf{\tilde{\Theta }}}}_{i}} \right) \right)}^{2}}}{4}\\ 
& +\frac{\left( \overset{\left( m \right)}{\mathop{{{\lambda }_{{{S}_{O}}\left( i \right)}}}}\,+\text{tr}\left( \left( {{\mathbf{A}}_{2,i}}-{{q}_{m,2.1}}{{\mathbf{A}}_{3,i}} \right)\overset{\left( m \right)}{\mathop{{{{\mathbf{\tilde{\Theta }}}}_{i}}}}\, \right) \right)}{2}\text{tr}\left( \left( {{\mathbf{A}}_{2,i}}-{{q}_{m,2.1}}{{\mathbf{A}}_{3,i}} \right)\left( {{{\mathbf{\tilde{\Theta }}}}_{i}}-\overset{\left( m \right)}{\mathop{{{{\mathbf{\tilde{\Theta }}}}_{i}}}}\, \right) \right)\\ 
& +\frac{\left( \overset{\left( m \right)}{\mathop{{{\lambda }_{{{S}_{O}}\left( i \right)}}}}\,+\text{tr}\left( \left( {{\mathbf{A}}_{2,i}}-{{q}_{m,2.1}}{{\mathbf{A}}_{3,i}} \right)\overset{\left( m \right)}{\mathop{{{{\mathbf{\tilde{\Theta }}}}_{i}}}}\, \right) \right)}{2}\left( {{\lambda }_{{{S}_{O}}\left( i \right)}}-\overset{\left( m \right)}{\mathop{{{\lambda }_{{{S}_{O}}\left( i \right)}}}}\, \right),
\end{split}
\end{equation}
where ${\mathop {{\lambda }_{{{S}_{O}}\left( i \right)}}\limits^{\left( m \right)} }$ and ${\mathop {{{\tilde{\mathbf{\Theta}} }_i}}\limits^{\left( m \right)} }$ are the given point in the $m$-th iteration. 
Therefore, problem (P5) can be efficiently solved by using CVX.

\section{Numerical Results}
In this section, we verify the effectiveness of our proposed algorithm in IOS-assisted ISAC system through numerical results. We consider a three-dimensional coordinate system in this section. The IOS and BS are located at (0, 0, 50m,) and (25m, 25m, 10m), respectively. ${{K}_{R}}=2$ user, ${{O}_{R}}=2$ sensed targets and are located randomly distributed in the reflective area whose coordinate is $\left( -100\le x\le 100,-100\le y\le 0,z=0 \right)$. ${{K}_{T}}=2$ users, ${{O}_{T}}=2$ sensed targets are located randomly distributed in the transmissive area whose coordinate are $\left( -100\le x\le 100,0\le y\le 100,z=0 \right)$. Other main system parameters are listed in Table I. We compare the proposed algorithm for ES, MS, and TS protocols with the other two benchmarks:

Conventional IRS (C-IRS): In this scheme, $N$ elements are evenly divided into two equal groups, which serve users and sensed targets in reflective and transmissive areas respectively, i.e., for elements serving the reflective area, the amplitude of their coefficients satisfies ${{\alpha }_{n,R}}=1,{{\alpha }_{n,T}}=0$ while for elements serving transmissive area, the amplitude of their coefficients satisfies ${{\alpha }_{n,R}}=0,{{\alpha }_{n,T}}=1$.

Uniform Energy Allocation (UEA): In this scheme, for each element, the amplitude of reflective and transmissive coefficients are equal. i.e., ${{\alpha }_{n,R}}={{\alpha }_{n,T}}={1}/{2}\;,\forall n$.
\begin{table}[t]
\centering
\small
\caption{Main Simulation Parameters}
\begin{tabular}{|c|c|c|c|c|c|}
\hline Parameter & Value & Parameter & Value & Parameter & Value\\
\hline${{N}_{t}}$ & $8$ & ${{N}_{r}}$ & $8$ & $N$ & $16$\\
\hline${{D}_{k}}$ & $2$ & ${{M}_{k}}$ & $4$ & ${{R}_{\text{th}}}$ & $1$\\
\hline${{P}_{max}}$ & $300w$ & $\sigma_{i}^{2}$ & $-60\text{dbm}$ & $\sigma _{k}^{2}$ & $-60\text{dbm}$\\
\hline
\end{tabular}
\end{table}

\begin{figure}[htbp]
\centering
\begin{minipage}[t]{0.48\textwidth}
\centering
\includegraphics[width=7cm]{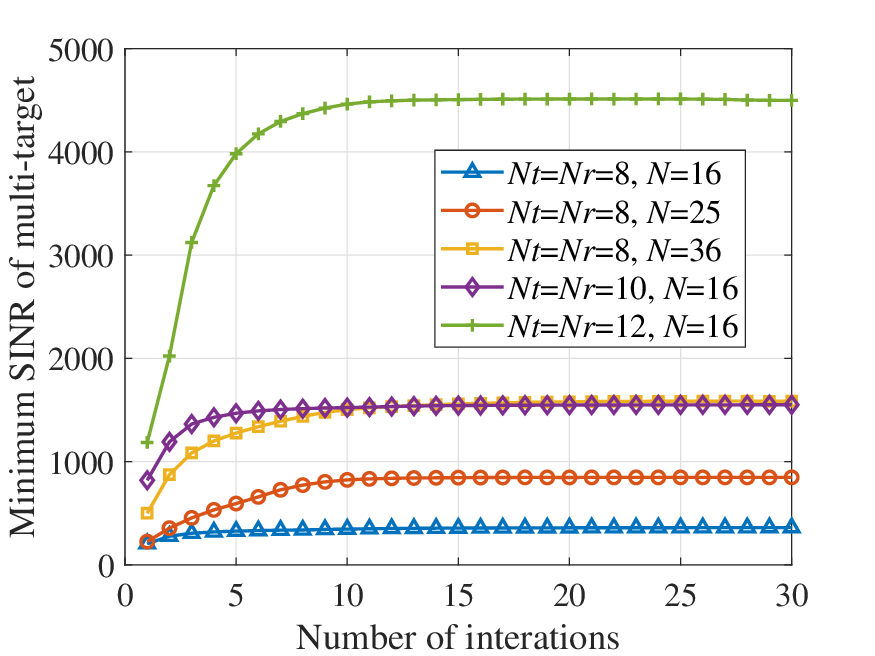}
\caption{Convergence behavior of the proposed robust joint optimization
algorithm.}
\end{minipage}
\begin{minipage}[t]{0.48\textwidth}
\centering
\includegraphics[width=7cm]{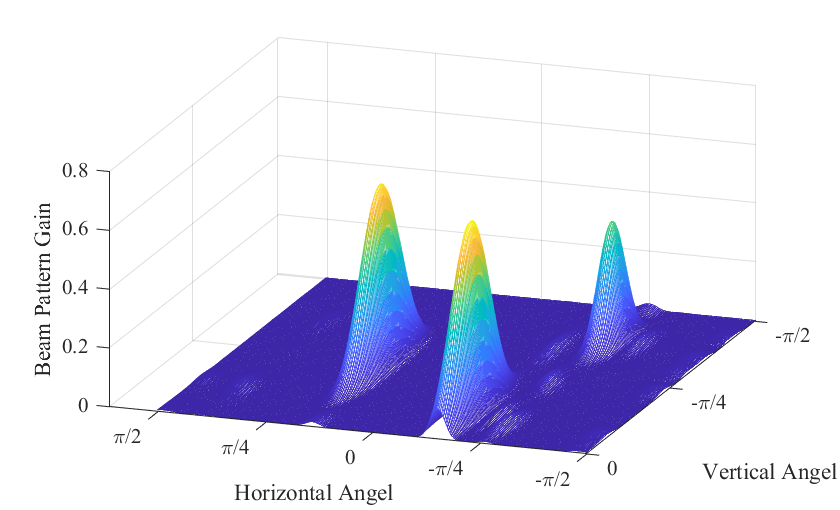}
\caption{3D beam pattern gain of the proposed algorithm.}
\end{minipage}
\end{figure}

First, the convergence of our algorithm is verified in IOS-enabled ISAC networks. Fig. 2 shows the convergence behavior of the proposed algorithm for ES protocol. It is obvious that the minimum SINR of the multi-target increases until convergence as the number of iterations increases, which verifies our proposed algorithm has good convergence. In addition, the number of transmitting/receiving antennas and elements has an impact on the complexity and convergence of the algorithm. Therefore, we set ${{N}_{t}}={{N}_{r}}=8,N=16$ as the benchmark, and change the value of ${N_t},{N_r},N$ and observe the robustness of the algorithm. It is depicted in the Fig. 2 that our algorithm can always converge in 20 iterations under different conditions. 

Fig. 3 shows the beam pattern gain versus horizontal and vertical angel in reflective area.  We assume that the horizontal and vertical angles of the three targets are ${{O}_{1}}=\left[ -0.5,-0.3 \right]$, ${{O}_{2}}=\left[ 0.5,-0.7 \right]$ and ${{O}_{3}}=\left[ 0.9,-1.2 \right]$ respectively. It’s obvious that our proposed beamforming scheme can make the beam gain pointing towards targets greater than other directions including the direction of the clutter, which indicates that our proposed algorithm can achieve accurate beamforming while ensuring the communication quality of users.
\begin{figure}[t]
\centering
\begin{minipage}[t]{0.48\textwidth}
\centering
\includegraphics[width=7cm]{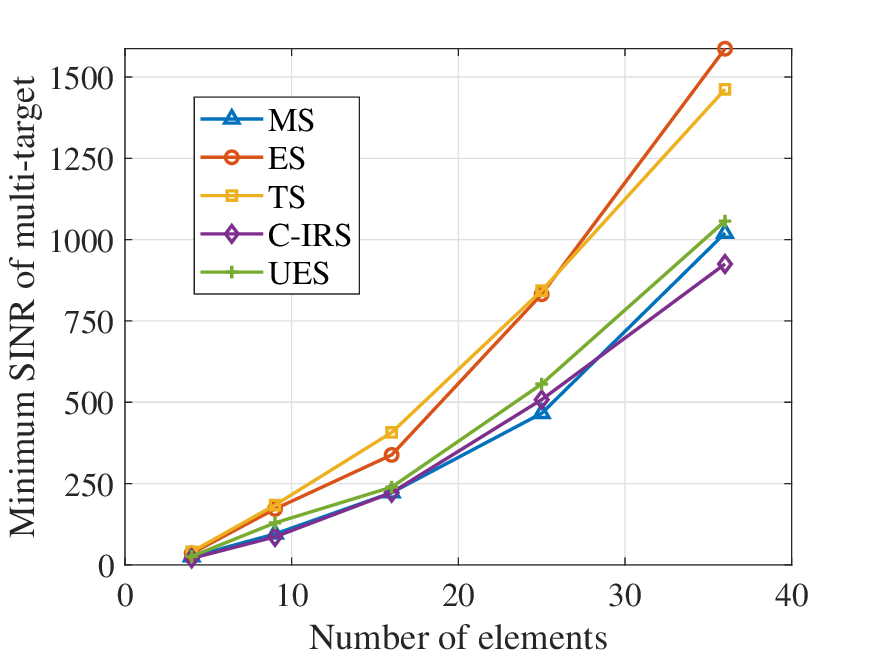}
\caption{SINR versus the number of elements.}
\end{minipage}
\begin{minipage}[t]{0.48\textwidth}
\centering
\includegraphics[width=7cm]{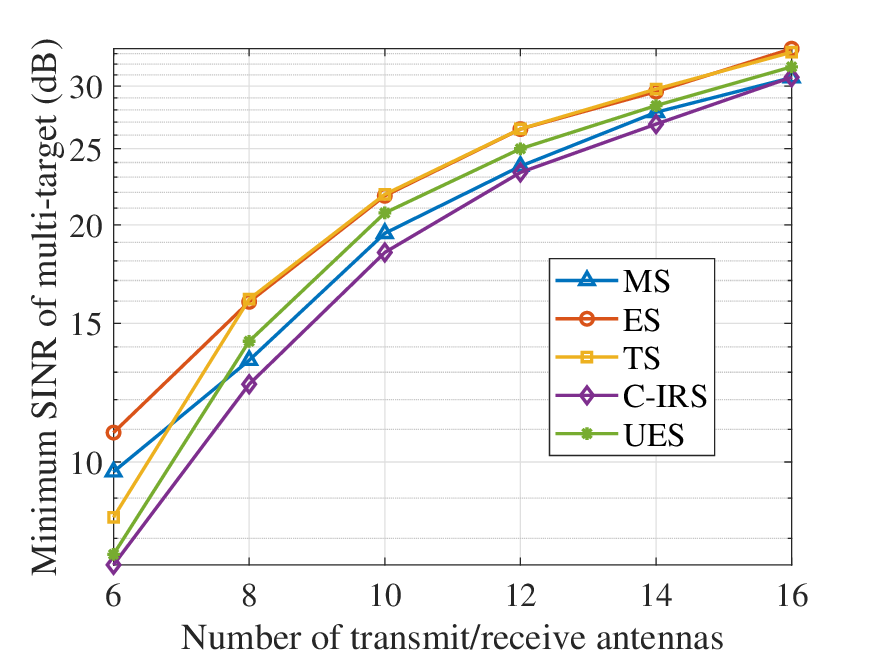}
\caption{SINR versus the number of transmit/receive antennas.}
\end{minipage}
\end{figure}

Then, we discuss the impact of the number of IOS elements on system performance. Fig. 4 shows the minimum SINR of multi-target versus the number of elements. It can be observed that the minimum SINR increases as the number of elements increases  because more components will provide greater spatial diversity gain. Specifically, ES and TS protocols are superior to the other three protocols due to the differences in design flexibility. ES and TS protocols exploit the ability of the whole elements that can reconstruct the amplitude and phase of the incident signal in the highest measure. The TS protocol can effectively eliminate interference in communication and improve the pertinence of sensing beams at the cost of time. However, when the number of elements is sufficient, the passive beamforming ability of IOS can effectively achieve the above-mentioned functions without sacrificing time. As a result, the ES protocol has the best performance when the number of elements exceeds 25. In general, increasing the number of IOS elements can improve system performance at a lower cost and has practical application value.

Next, Fig. 5 illustrates the variation of the minimum SINR of multi-target with the number of transmit/receive antennas. Obviously, increasing the number of antennas can greatly improve system performance without the need for additional transmission power overhead. More antennas can enhance active beamforming and increase beaming gain. UES and MS protocols cannot split the energy of incident signals into reflective and transmissive areas, resulting in a performance loss of approximately 4dB compared to the ES protocol. The performance of C-IRS is the worst mainly because the beam gain achieved by IOS is versus $N^2$ with increasing $N$. When IOS is divided into two groups equally, the beam gain in each area is reduced to one-quarter of the original. Although two groups of IRS serving different areas can reduce interference, it cannot compensate for the performance loss caused by the decrease in the number of IRS elements.
\begin{figure}[t]
\centering
\begin{minipage}[t]{0.48\textwidth}
\centering
\includegraphics[width=7cm]{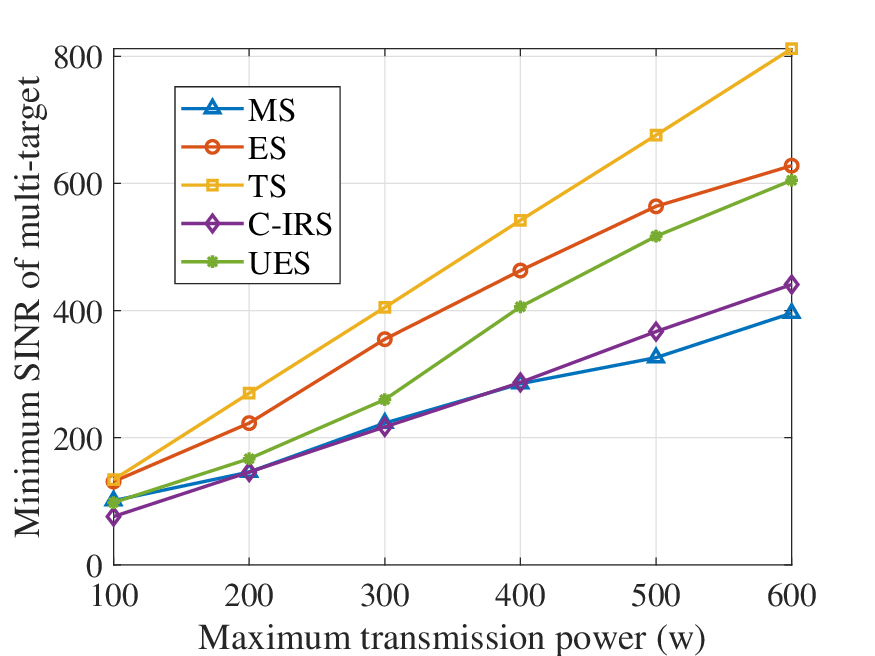}
\caption{SINR versus the maximum transmit power.}
\end{minipage}
\begin{minipage}[t]{0.48\textwidth}
\centering
\includegraphics[width=7cm]{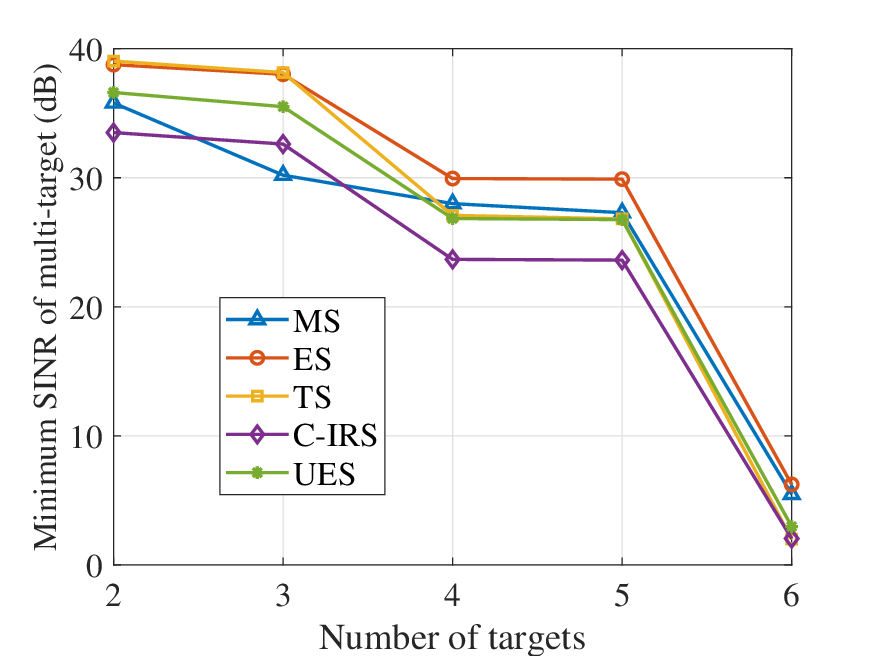}
\caption{SINR versus the number of targets.}
\end{minipage}
\end{figure}
Fig. 6 depicts the minimum SINR of multi-target sensing versus the maximum transmit power of BS. It can be seen that greater transmission power can achieve better system performance. However, the performance improvement brought about by increasing transmit power is not as significant as increasing the number of antennas and IRS elements. When the transmission power reaches a certain threshold, the SINR of the target’s echo signal is mainly limited by the interference brought by other targets and clutters. Increasing the number of antennas or IOS elements can achieve more accurate beamforming and reduce interference while increasing the transmission power can only reduce the interference caused by noise significantly. Although the performance of TS protocol is slightly better than ES protocol as the number of antennas increases, the TS protocol requires synchronization of the switching cycles between the BS and IRS, which will be a challenge in hardware implementation.

Finally, we investigate the relationship between the minimum SINR of multi-target sensing and the number of targets in Fig. Here, we assume ${{O}_{R}}=\left\lceil {O}/{2}\; \right\rceil ,{{O}_{T}}=\left\lfloor {O}/{2}\; \right\rfloor$. It is obvious that the minimum SINR of multi-target decreases as the increase of the number of targets. The location of newly added targets will affect the extent of system performance degradation. Specifically, when the location of a new target is close to the location of an original target, the channel correlation between these two targets is strong, which makes it difficult to distinguish the echo signals from these two targets. Conversely, if the channels of the new target and the original targets have good orthogonality, radar signals can achieve good spatial multiplexing for multi-target sensing, which makes system performance almost unchanged. 
\section{Conclusions}
In this paper, we investigate the minimum SINR of multi-target sensing maximization problem for IOS-assisted ISAC system, where IOS enabled dual-function BS to achieve 360-degree ISAC coverage. Specifically, radar receive vectors, IOS reflective, and transmissive coefficient and 
beamforming matrices are jointly designed while satisfying the requirements for reachable rates of multi-user. Because of the high coupling between optimization variables, the primal problem is decomposed into three sub-problems based on the introduction of an auxiliary variable and BCD algorithm. To solve the sub-problem, SDR and SCA methods are utilized to convert the non-convex objective functions and constraints into convex functions. Then, the iterative rank minimization and penalty function method are used to ensure the equivalence of the conversion. In addition, We extend the algorithm design to MS and TS protocols to adapt to different scenarios. Finally, the simulation results validate the effectiveness of the proposed algorithm from multiple perspectives, including performance improvement compared with other benchmark schemes under different parameters, beam pattern gain and convergence.

\appendices
\section{Proof of Proposition 1}
${{q}_{m,x}}$ is a function with respect to $\left( {{\mathbf{g}}_{i}},{{\mathbf{W}}_{k}},{{\mathbf{M}}_{t}},{{\mathbf{\theta }}_{R}},{{\mathbf{\theta }}_{T}} \right)$. Because the proposed algorithm is based on BCD, we only consider changes in the corresponding optimization variables in each step of our proof.

(1) ${{q}_{m,2.1}}\left( \mathbf{\tilde{W}}_{k}^{\left( m,2.2 \right)},\mathbf{\tilde{M}}_{t}^{\left( m,2.2 \right)} \right)\ge {{q}_{m,1}^{*}}$

The proof is based on contradiction, i.e., we assume ${{q}_{m,1}^{*}}>{{q}_{m,2.1}}\left( \mathbf{\tilde{W}}_{k}^{\left( m,2.2 \right)},\mathbf{\tilde{M}}_{t}^{\left( m,2.2 \right)} \right)$. When problem (P1.1) is solved in $m$-th, ${q_{m,1}}$ can be obtained and $\left( \mathbf{\tilde{W}}_{k}^{\left( m,1 \right)},\mathbf{\tilde{M}}_{t}^{\left( m,1 \right)} \right)$ express a feasible solution. Therefore,  $\left( \mathbf{\tilde{W}}_{k}^{\left( m,1 \right)},\mathbf{\tilde{M}}_{t}^{\left( m,1 \right)} \right)$ can also be a feasible solution for problem (P2.2) and we can get
\begin{equation}\label{ap1}
{{q}_{m,1}}={{q}_{m,2.2}}\left( \mathbf{\tilde{W}}_{k}^{\left( m,1 \right)},\mathbf{\tilde{M}}_{t}^{\left( m,1 \right)} \right).
\end{equation}

When problem (P2.2) is solved in $m$-th iteration, optimal solution $\left( \mathbf{\tilde{W}}_{k}^{\left( m,2.2 \right)},\mathbf{\tilde{M}}_{t}^{\left( m,2.2 \right)} \right)$ and optimal value ${{q}_{m,2.2}}\left( \mathbf{\tilde{W}}_{k}^{\left( m,2.2 \right)},\mathbf{\tilde{M}}_{t}^{\left( m,2.2 \right)} \right)$ can be obtained. Moreover, $\left( \mathbf{\tilde{W}}_{k}^{\left( m,2.2 \right)},\mathbf{\tilde{M}}_{t}^{\left( m,2.2 \right)} \right)$ can also be a feasible solution for problem (P2.1) and there must exist a target $i'$ that satisfies 
\begin{equation}
i'=\underset{i}{\mathop{\arg \min }}\,q_{m,2.1}^{\left( i \right)}\left( \mathbf{\tilde{W}}_{k}^{\left( m,2.2 \right)},\mathbf{\tilde{M}}_{t}^{\left(m,2.2 \right)} \right),
\end{equation}
where $q_{m,2.1}^{\left( i \right)}$ represents the SINR of $i$-th target in $\left( P2.1 \right)$, which can be expressed as
\begin{equation}\label{ap2}
q_{m,2.1}^{\left( i \right)}=\frac{\text{tr}\left( {{{\tilde{\mathbf{S}}}}_{i,i}}\left( \sum\limits_{k=1}^{K}{\tilde{\mathbf{W}}_{k}}+\sum\limits_{t=1}^{{{N}_{T}}}{\tilde{\mathbf{M}}_{t}} \right) \right)}{\text{tr}\left( \sum\limits_{o=1+\delta \left( {i - {O_R}} \right)\cdot{O_R},o\ne i}^{{{O}_{R}+\delta \left( {i - {O_R}} \right)\cdot{O_T}}}{{{{\tilde{\mathbf{S}}}}_{i,o}}}\left( \sum\limits_{k=1}^{K}{\tilde{\mathbf{W}}_{k}}+\sum\limits_{t=1}^{{{N}_{T}}}{\tilde{\mathbf{M}}_{t}} \right) \right)+{{c}_{i}}}.
\end{equation}
Next, $q_{m,2.2}^{\left( i \right)}$ can be expressed as
\begin{equation}\label{ap3}
{q_{m,2.2}^{\left( i \right)}}=\frac{\text{tr}\left( \left( {{{\tilde{\mathbf{S}}}}_{i,i}}-{{q}_{m,1}}\sum\limits_{o=1+\delta \left( {i - {O_R}} \right)\cdot{O_R},o\ne i}^{{{O}_{R}+\delta \left( {i - {O_R}} \right)\cdot{O_T}}}{{{{\tilde{\mathbf{S}}}}_{i,o}}} \right)\left( \sum\limits_{k=1}^{K}{\tilde{\mathbf{W}}_{k}}+\sum\limits_{t=1}^{{{N}_{T}}}{\tilde{\mathbf{M}}_{t}} \right) \right)}{{{c}_{i}}}.
\end{equation}
The numerical relationship between $q_{m,2.1}^{\left( i \right)}$ and $q_{m,2.2}^{\left( i \right)}$ can be expressed as
\begin{equation}\label{ap4}
\begin{split}
  & q_{m,2.2}^{\left( i \right)}=q_{m,2.1}^{\left( i \right)} \\ 
 & +\frac{\text{tr}\left( \left( \sum\limits_{o=1+\delta \left( {i - {O_R}} \right)\cdot{O_R},o\ne i}^{{{O}_{R}+\delta \left( {i - {O_R}} \right)\cdot{O_T}}}{{{{\tilde{\mathbf{S}}}}_{i,o}}} \right)\left( \sum\limits_{k=1}^{K}{\tilde{\mathbf{W}}_{k}}+\sum\limits_{t=1}^{{{N}_{T}}}{\tilde{\mathbf{M}}_{t}}\right) \right)}{{{c}_{i}}}\cdot\left( q_{m,2.1}^{\left( i \right)}-q_{m,1}\right),
\end{split}
\end{equation}
It's obvious that if $q_{m,1}>q_{m,2.1}^{\left( i \right)}\left( \tilde{\mathbf{W}}_{k}^{\left( m,1 \right)},\tilde{\mathbf{M}}_{t}^{\left( m,1 \right)} \right)$, $q_{m,2.1}^{\left( i \right)}>q_{m,2.2}^{\left( i \right)}$. Then we can get 
\begin{equation}\label{ap5}
\begin{split}
& {{q}_{m,1}}\overset{\left( a \right)}{\mathop{>}}\,q_{m,2.1}\left( \mathbf{\tilde{W}}_{k}^{\left( m,2.2 \right)},\mathbf{\tilde{M}}_{t}^{\left( m,2.2 \right)} \right)\overset{\left( b \right)}{\mathop{=}}\,{{q}_{m,2.1}^{\left( i' \right)}}\left( \mathbf{\tilde{W}}_{k}^{\left( m,2.2 \right)},\mathbf{\tilde{M}}_{t}^{\left( m,2.2 \right)} \right) \\ 
& \overset{\left( c \right)}{\mathop{>}}\,q_{m,2.2}^{\left( i' \right)}\left( \mathbf{\tilde{W}}_{k}^{\left( m,2.2 \right)},\mathbf{\tilde{M}}_{t}^{\left( m,2.2 \right)} \right)\overset{\left( d \right)}{\mathop{\ge }}\,{{q}_{m,2.2}}\left( \mathbf{\tilde{W}}_{k}^{\left( m,2.2 \right)},\mathbf{\tilde{M}}_{t}^{\left( m,2.2 \right)} \right).
\end{split}
\end{equation}
The procedure $(a)$ is due to assumptions in the contradiction, $(b)$ is due to the definition of $i'$, $(c)$ is due to Eq. (\ref{ap4}) and $(d)$ is due to the definition of ${{q}_{m,2.2}}$. Recalling Eq. (\ref{ap1}), we have
\begin{equation}
{{q}_{m,1}}={{q}_{m,2.2}}\left( \mathbf{\tilde{W}}_{k}^{\left( m,1 \right)},\mathbf{\tilde{M}}_{t}^{\left( m,1 \right)} \right)>{{q}_{m,2.2}}\left( \mathbf{\tilde{W}}_{k}^{\left( m,2.2 \right)},\mathbf{\tilde{M}}_{t}^{\left( m,2.2 \right)} \right).
\end{equation}
However, problem (P2.2) is a convex problem while $\left( {{\bf{\tilde W}}_k^{\left( {m,2.2} \right)},{\bf{\tilde M}}_t^{\left( {m,2.2} \right)}} \right)$ is an optimal solution and $\left( {{\bf{\tilde W}}_k^{\left( {m,1} \right)},{\bf{\tilde M}}_t^{\left( {m,1} \right)}} \right)$ is a feasible solution, which is contradictory and the proof is completed.

(2)  ${{q}_{m,3.1}}\left( \mathbf{\theta }_{R}^{\left( m,3.1 \right)},\mathbf{\theta }_{T}^{\left( m,3.1 \right)} \right)\ge {{q}_{m,2.1}}\left( \mathbf{\theta }_{R}^{\left( m,2.2 \right)},\mathbf{\theta }_{T}^{\left( m,2.2 \right)} \right)$

The proof process is similar to proof ${{q}_{m,2.1}}\left( \mathbf{W}_{k}^{\left( m,2.2 \right)},\mathbf{M}_{t}^{\left( m,2.2 \right)} \right)\ge q_{m,1}^{*}$, and we omit this process here.

(3) ${{q}_{m+1,1}}\left( \mathbf{g}_{i}^{\left( m+1,1 \right)} \right)\ge {{q}_{m,3.1}}\left( \mathbf{g}_{i}^{\left( m,1 \right)} \right)$

The change of ${{\mathbf{g}}_{i}}$ will not affect the feasibility of the original problem (P0). $\mathbf{g}_{i}^{\left( m+1,1 \right)}$ is the optimal solution for maximizing $\text{SIN}{{\text{R}}_{i}}$ given $\left( {{\mathbf{W}}_{k}},{{\mathbf{M}}_{t}},{{\mathbf{\theta }}_{R}},{{\mathbf{\theta }}_{T}} \right)$ after the completion of the $m$-th iteration. Therefore,
\begin{equation}
q_{m+1,1}^{\left( i \right)}\left( \mathbf{g}_{i}^{\left( m+1,1 \right)} \right)\ge q_{m,3.1}^{\left( i \right)}\left( \mathbf{g}_{i}^{\left( m,1 \right)} \right),\forall i,
\end{equation}
Then we have
\begin{equation}
{{q}_{m+1,1}}=\underset{i}{\mathop{\min }}\,\left( q_{m+1,1}^{\left( i \right)}\left( \mathbf{g}_{i}^{\left( m+1,1 \right)} \right) \right)\ge {{q}_{m,3.1}}=\underset{i}{\mathop{\min }}\,\left( q_{m,3.1}^{\left( i \right)}\left( \mathbf{g}_{i}^{\left( m,1 \right)} \right) \right).
\end{equation}
and the proof is completed.
\section{Proof of Proposition 3}
First, we prove $\log \det \left(\mathbf{V}_{k}^{-1} \right)={{R}_{k}}$, ${{\mathbf{E}}_{k}}$ can be expressed as
\begin{equation}
{{\bf{E}}_k} = \left( {{{\bf{I}}_{{D_k}}} - {{\bf{G}}_k}{{\bf{H}}_k}{{\bf{W}}_k}} \right){\left( {{{\bf{I}}_{{D_k}}} - {{\bf{G}}_k}{{\bf{H}}_k}{{\bf{W}}_k}} \right)^H}.
\end{equation}
By plugging back this optimum value of  ${{{\bf{G}}_k}}$ in Eq. (\ref{gkvk}) and using Woodbury matrix identity, 
\begin{equation}
{{\mathbf{E}}_{k}}=\left( {{\mathbf{I}}_{{{D}_{k}}}}-\mathbf{W}_{k}^{H}\mathbf{H}_{k}^{H}{{\left( \sigma _{k}^{2}{{\mathbf{I}}_{{{M}_{k}}}}+\sum\limits_{k=1}^{K}{{{\mathbf{H}}_{k}}{{\mathbf{W}}_{l}}\mathbf{W}_{l}^{H}\mathbf{H}_{k}^{H}}+\sum\limits_{t=1}^{{{N}_{T}}}{{{\mathbf{H}}_{k}}{{\mathbf{m}}_{t}}\mathbf{m}_{t}^{H}\mathbf{H}_{k}^{H}} \right)}^{-1}}{{\mathbf{H}}_{k}}{{\mathbf{W}}_{k}} \right).
\end{equation}
Then $\log \det \left( \mathbf{E}_{k}^{-1} \right)$ can be expressed as
\begin{equation}
\log \det \left( \mathbf{E}_{k}^{-1} \right)=\log \det \left( {{\mathbf{I}}_{{{D}_{k}}}}+\mathbf{W}_{k}^{H}\mathbf{H}_{k}^{H}\mathbf{J}_{k}^{-1}{{\mathbf{H}}_{k}}{{\mathbf{W}}_{k}} \right)={{R}_{k}}.
\end{equation}
Next, $\text{tr}\left( {{\mathbf{V}_k}{\mathbf{E}_k}} \right) - \log \det \left( {{\mathbf{V}_k}} \right)$ and ${R_k}$ are convex function and concave function on ${{\mathbf{\Theta }}_{R}},{{\mathbf{\Theta }}_{T}}$ respectively. Therefore, when ${{\mathbf{V}}_{k}}=\mathbf{E}_{k}^{-1}$, $\text{tr}\left( {{\mathbf{V}}_{k}}{{\mathbf{E}}_{k}} \right)={{D}_{k}}$ and ${{R}_{k}}\ge {{R}_{\text{th}}}$ is equivalent to $\text{tr}\left( {{\mathbf{V}}_{k}}{{\mathbf{E}}_{k}} \right)-\log \det \left( {{\mathbf{V}}_{k}} \right)\le {{D}_{k}}-{{R}_{\text{th}}}$.
\bibliographystyle{IEEEtran}
\bibliography{reference}

\begin{thebibliography}{10}
\providecommand{\url}[1]{#1}
\csname url@samestyle\endcsname
\providecommand{\newblock}{\relax}
\providecommand{\bibinfo}[2]{#2}
\providecommand{\BIBentrySTDinterwordspacing}{\spaceskip=0pt\relax}
\providecommand{\BIBentryALTinterwordstretchfactor}{4}
\providecommand{\BIBentryALTinterwordspacing}{\spaceskip=\fontdimen2\font plus
\BIBentryALTinterwordstretchfactor\fontdimen3\font minus
  \fontdimen4\font\relax}
\providecommand{\BIBforeignlanguage}[2]{{%
\expandafter\ifx\csname l@#1\endcsname\relax
\typeout{** WARNING: IEEEtran.bst: No hyphenation pattern has been}%
\typeout{** loaded for the language `#1'. Using the pattern for}%
\typeout{** the default language instead.}%
\else
\language=\csname l@#1\endcsname
\fi
#2}}
\providecommand{\BIBdecl}{\relax}
\BIBdecl

\bibitem{10077112}
X.~Li, Y.~Gong, K.~Huang, and Z.~Niu, ``Over-the-air integrated sensing,
  communication, and computation in {IoT} networks,'' \emph{IEEE Wireless
  Commun.}, vol.~30, no.~1, pp. 32--38, Feb. 2023.

\bibitem{9737357}
F.~Liu, Y.~Cui, C.~Masouros, J.~Xu, T.~X. Han, Y.~C. Eldar, and S.~Buzzi,
  ``Integrated sensing and communications: Toward dual-functional wireless
  networks for 6{G} and beyond,'' \emph{IEEE J. Sel. Areas Commun.}, vol.~40,
  no.~6, pp. 1728--1767, Mar. 2022.

\bibitem{7279172}
A.~R. Chiriyath, B.~Paul, G.~M. Jacyna, and D.~W. Bliss, ``Inner bounds on
  performance of radar and communications co-existence,'' \emph{IEEE Trans.
  Signal Process.}, vol.~64, no.~2, pp. 464--474, Sept. 2016.

\bibitem{10001144}
Y.~Xiong, F.~Liu, Y.~Cui, W.~Yuan, and T.~X. Han, ``Flowing the information
  from {Shannon} to {Fisher}: Towards the fundamental tradeoff in {ISAC},'' in
  \emph{Proc. IEEE Global Commun. Conf. (GLOBECOM)}, 2022, pp. 5601--5606.

\bibitem{4268440}
G.~N. Saddik, R.~S. Singh, and E.~R. Brown, ``Ultra-wideband multifunctional
  communications/radar system,'' \emph{IEEE Trans. Microwave Theory Tech.},
  vol.~55, no.~7, pp. 1431--1437, Jul. 2007.

\bibitem{9724187}
Z.~Xiao and Y.~Zeng, ``Waveform design and performance analysis for full-duplex
  integrated sensing and communication,'' \emph{IEEE J. Sel. Areas Commun.},
  vol.~40, no.~6, pp. 1823--1837, Mar. 2022.

\bibitem{7347464}
A.~Hassanien, M.~G. Amin, Y.~D. Zhang, and F.~Ahmad, ``Dual-function
  radar-communications: Information embedding using sidelobe control and
  waveform diversity,'' \emph{IEEE Trans. Signal Process.}, vol.~64, no.~8, pp.
  2168--2181, Dec. 2016.

\bibitem{9529026}
M.~F. Keskin, H.~Wymeersch, and V.~Koivunen, ``{MIMO-OFDM} joint
  radar-communications: Is {ICI} friend or foe?'' \emph{IEEE J. Sel. Top. Sign.
  Proces.}, vol.~15, no.~6, pp. 1393--1408, Sept. 2021.

\bibitem{9829188}
H.~Lin and J.~Yuan, ``Orthogonal delay-doppler division multiplexing
  modulation,'' \emph{IEEE Trans. Wireless Commun.}, vol.~21, no.~12, pp.
  11\,024--11\,037, Jul. 2022.

\bibitem{9838684}
H.~Lin and J.~Yuan, ``Multicarrier modulation on delay-doppler plane: Achieving
  orthogonality with fine resolutions,'' in \emph{Proc. IEEE Int. Conf. Commun.
  (ICC)}, 2022, pp. 2417--2422.

\bibitem{9728752}
Q.~Zhang, H.~Sun, X.~Gao, X.~Wang, and Z.~Feng, ``Time-division {ISAC} enabled
  connected automated vehicles cooperation algorithm design and performance
  evaluation,'' \emph{IEEE J. Sel. Areas Commun.}, vol.~40, no.~7, pp.
  2206--2218, Mar. 2022.

\bibitem{8094973}
C.~Shi, F.~Wang, M.~Sellathurai, J.~Zhou, and S.~Salous, ``Power
  minimization-based robust {OFDM} radar waveform design for radar and
  communication systems in coexistence,'' \emph{IEEE Trans. Signal Process.},
  vol.~66, no.~5, pp. 1316--1330, Nov. 2018.

\bibitem{9359665}
X.~Chen, Z.~Feng, Z.~Wei, P.~Zhang, and X.~Yuan, ``Code-division {OFDM} joint
  communication and sensing system for 6{G} machine-type communication,''
  \emph{IEEE Internet Things J.}, vol.~8, no.~15, pp. 12\,093--12\,105, Feb.
  2021.

\bibitem{8288677}
F.~Liu, C.~Masouros, A.~Li, H.~Sun, and L.~Hanzo, ``{MU-MIMO} communications
  with {MIMO} radar: From co-existence to joint transmission,'' \emph{IEEE
  Trans. Wireless Commun.}, vol.~17, no.~4, pp. 2755--2770, Feb. 2018.

\bibitem{9724205}
X.~Liu, T.~Huang, and Y.~Liu, ``Transmit design for joint {MIMO} radar and
  multiuser communications with transmit covariance constraint,'' \emph{IEEE J.
  Sel. Areas Commun.}, vol.~40, no.~6, pp. 1932--1950, Mar. 2022.

\bibitem{9086766}
M.~A. ElMossallamy, H.~Zhang, L.~Song, K.~G. Seddik, Z.~Han, and G.~Y. Li,
  ``Reconfigurable intelligent surfaces for wireless communications:
  Principles, challenges, and opportunities,'' \emph{IEEE Trans. Cognitive
  Commun. and Networking}, vol.~6, no.~3, pp. 990--1002, May 2020.

\bibitem{9361184}
W.~Lu, Q.~Lin, N.~Song, Q.~Fang, X.~Hua, and B.~Deng, ``Target detection in
  intelligent reflecting surface aided distributed mimo radar systems,''
  \emph{IEEE Sensors Lett.}, vol.~5, no.~3, pp. 1--4, Feb. 2021.

\bibitem{9456027}
H.~Zhang, H.~Zhang, B.~Di, K.~Bian, Z.~Han, and L.~Song, ``Metalocalization:
  Reconfigurable intelligent surface aided multi-user wireless indoor
  localization,'' \emph{IEEE Trans. Wireless Commun.}, vol.~20, no.~12, pp.
  7743--7757, Jun. 2021.

\bibitem{9724202}
X.~Shao, C.~You, W.~Ma, X.~Chen, and R.~Zhang, ``Target sensing with
  intelligent reflecting surface: Architecture and performance,'' \emph{IEEE J.
  Sel. Areas Commun.}, vol.~40, no.~7, pp. 2070--2084, Mar. 2022.

\bibitem{9938373}
K.~Meng, Q.~Wu, R.~Schober, and W.~Chen, ``Intelligent reflecting surface
  enabled multi-target sensing,'' \emph{IEEE Trans. Commun.}, vol.~70, no.~12,
  pp. 8313--8330, Nov. 2022.

\bibitem{9705498}
A.~Liu, Z.~Huang, M.~Li, Y.~Wan, W.~Li, T.~X. Han, C.~Liu, R.~Du, D.~K.~P. Tan,
  J.~Lu, Y.~Shen, F.~Colone, and K.~Chetty, ``A survey on fundamental limits of
  integrated sensing and communication,'' \emph{IEEE Commun. Surveys Tuts.},
  vol.~24, no.~2, pp. 994--1034, Feb. 2022.

\bibitem{9416177}
X.~Wang, Z.~Fei, Z.~Zheng, and J.~Guo, ``Joint waveform design and passive
  beamforming for {RIS}-assisted dual-functional radar-communication system,''
  \emph{IEEE Trans. Veh. Technol.}, vol.~70, no.~5, pp. 5131--5136, Apr. 2021.

\bibitem{9729741}
Y.~He, Y.~Cai, H.~Mao, and G.~Yu, ``{RIS}-assisted communication radar
  coexistence: Joint beamforming design and analysis,'' \emph{IEEE J. Sel.
  Areas Commun.}, vol.~40, no.~7, pp. 2131--2145, Mar. 2022.

\bibitem{10131933}
K.~Meng, Q.~Wu, W.~Chen, E.~Paolini, and E.~Matricardi, ``Intelligent surface
  empowered sensing and communication: A novel mutual assistance design,''
  \emph{IEEE Commun. Lett.}, pp. 1--1, May 2023.

\bibitem{10143420}
M.~Hua, Q.~Wu, W.~Chen, O.~A. Dobre, and A.~Lee~Swindlehurst, ``Secure
  intelligent reflecting surface aided integrated sensing and communication,''
  \emph{IEEE Trans. Wireless Commun.}, pp. 1--1, Jun. 2023.

\bibitem{9424177}
Y.~Liu, X.~Liu, X.~Mu, T.~Hou, J.~Xu, M.~Di~Renzo, and N.~Al-Dhahir,
  ``Reconfigurable intelligent surfaces: Principles and opportunities,''
  \emph{IEEE Commun. Surveys Tuts.}, vol.~23, no.~3, pp. 1546--1577, May 2021.

\bibitem{9570775}
Z.~Li, W.~Chen, and H.~Cao, ``Beamforming design and power allocation for
  transmissive rms-based transmitter architectures,'' \emph{IEEE Wireless
  Commun.Lett.}, vol.~11, no.~1, pp. 53--57, Oct. 2022.

\bibitem{9365009}
S.~Zeng, H.~Zhang, B.~Di, Y.~Tan, Z.~Han, H.~V. Poor, and L.~Song,
  ``Reconfigurable intelligent surfaces in 6{G}: Reflective, transmissive, or
  both?'' \emph{IEEE Commun. Lett.}, vol.~25, no.~6, pp. 2063--2067, Feb. 2021.

\bibitem{9690478}
Y.~Liu, X.~Mu, J.~Xu, R.~Schober, Y.~Hao, H.~V. Poor, and L.~Hanzo, ``{STAR}:
  Simultaneous transmission and reflection for 360 coverage by intelligent
  surfaces,'' \emph{IEEE Wireless Commun.}, vol.~28, no.~6, pp. 102--109, Dec.
  2021.

\bibitem{9570143}
X.~Mu, Y.~Liu, L.~Guo, J.~Lin, and R.~Schober, ``Simultaneously transmitting
  and reflecting {(STAR)} {RIS} aided wireless communications,'' \emph{IEEE
  Trans. Wireless Commun.}, vol.~21, no.~5, pp. 3083--3098, Oct. 2022.

\bibitem{10050406}
Z.~Wang, X.~Mu, and Y.~Liu, ``{STARS} enabled integrated sensing and
  communications,'' \emph{IEEE Trans. Wireless Commun.}, pp. 1--1, Feb. 2023.

\bibitem{9652071}
F.~Liu, Y.-F. Liu, A.~Li, C.~Masouros, and Y.~C. Eldar, ``Cramér-rao bound
  optimization for joint radar-communication beamforming,'' \emph{IEEE Trans.
  Signal Process.}, vol.~70, pp. 240--253, Dec. 2022.

\bibitem{9163260}
J.~Zhou, H.~Li, and W.~Cui, ``Low-complexity joint transmit and receive
  beamforming for mimo radar with multi-targets,'' \emph{IEEE Signal Processing
  Lett.}, vol.~27, pp. 1410--1414, Aug. 2020.

\bibitem{sun2018rank}
C.~Sun, ``Rank-constrained optimization: Algorithms and applications,'' Ph.D.
  dissertation, The Ohio State University, 2018.

\bibitem{5756489}
Q.~Shi, M.~Razaviyayn, Z.-Q. Luo, and C.~He, ``An iteratively weighted {MMSE}
  approach to distributed sum-utility maximization for a {MIMO} interfering
  broadcast channel,'' \emph{IEEE Trans. Signal Process.}, vol.~59, no.~9, pp.
  4331--4340, Feb. 2011.

\end{thebibliography}
\end{document}